\documentclass{aa}  

\usepackage{graphicx}
\usepackage{enumitem}
\usepackage{txfonts}
\begin{document}

   \title{Determination of the escape velocity of the Milky Way using a proper motion selected halo sample}
   \titlerunning{Determination of the escape velocity}
      \author{Helmer H. Koppelman \inst{1,2}
          \and
          Amina Helmi \inst{1}
          }

        \institute{
        Kapteyn Astronomical Institute, University of Groningen, Landleven 12, 9747 AD Groningen, The Netherlands \and 
        School of Natural Sciences, Institute for Advanced Study, 1 Einstein Drive, Princeton, NJ 08540, USA\\
              \email{koppelman@ias.edu}}

   \date{submitted June 28, 2020}

  \abstract
   {The {\it Gaia} mission has provided the largest catalogue ever of sources with tangential velocity information. However, using this catalogue for dynamical studies is difficult because most of the stars lack line-of-sight velocity measurements. Recently, we presented a selection of $\sim 10^7$ halo stars with accurate distances that have been selected based on their photometry and proper motions.}
   {Using this sample, we model the tail of the velocity distribution in the stellar halo, locally and as a function of distance. Our goal is to measure the escape velocity, and herewith to constrain the mass of our Galaxy.}
   {We fit the tail of the velocity distribution with a power-law distribution, a commonly used approach first established by \cite{Leonard1990THESPEED}. For the first time ever we use tangential velocities measured accurately for an unprecedented number of halo stars to estimate the escape velocity.}
   {In the solar neighbourhood, we obtain a very precise estimate of the escape velocity which is $497^{+8}_{-8}~{\rm km/s}$. This estimate is most likely biased low, our best guess is by 10\%. As a result, the true escape velocity most likely is closer to $550~{\rm km/s}$.
   The escape velocity directly constrains the total mass of the Milky Way. To find the best fitting halo mass and concentration parameter we adjusted the dark (spherical NFW) halo of a realistic Milky Way potential while keeping the circular velocity at the solar radius fixed at $v_c(R_\odot) = 232.8~{\rm km/s}$. The resulting halo parameters are $M_{200}^{+10\%} = 1.11^{+0.08}_{-0.07} \cdot10^{12} ~{\rm M}_\odot$ and concentration parameter $c^{+10\%} = 11.8^{+0.3}_{-0.3}$, where we use the explicit notation to indicate that these have been corrected for the 10\% bias.
   The slope of the escape velocity with galactocentric distance is as expected in the inner Galaxy based on Milky Way models. Curiously, we find a disagreement beyond the solar radius where the estimated escape velocity is larger than at the solar radius. This result is likely an effect of a change in the shape of the velocity distribution and could be related to the presence of velocity clumps. A tentative analysis of the escape velocity as a function of $(R,z)$ shows that the slope is shallower than expected for a spherical halo using standard values for the characteristic parameters describing the galactic disc.
   }
   {}

   \keywords{Galaxy: kinematics and dynamics --
             Galaxy: structure --
             Galaxy: fundamental parameters}

   \maketitle
%
\section{Introduction}
Numerous studies have attempted to measure the mass of the Milky Way,
yet it has been notoriously difficult to obtain precise and model
independent constraints. Most works now agree that the mass of the
Milky Way's dark matter halo is $10^{12}~{\rm M}_\odot$ within a
factor of two \citep[see Fig. 7 of][for a recent
compilation]{Callingham2019TheDynamics}. The kinematics of globular
clusters, dwarf galaxies, and halo stars have often been used in such
studies \citep{Kochanek1996TheWay, Xue2008, Watkins2010TheGalaxies, Deason2012TheHalo, Fragione2016, Posti2019MassHubble, Callingham2019TheDynamics, Fritz2020}. 
The timing argument and the properties of 
debris streams such as those from the Sagittarius dwarf \citep[e.g.][]{Dierickx2017a, Zaritsky2019ASurvey} have provided additional, yet similar constraints. In this work, we aim to derive a very precise estimate of the escape
velocity near the Sun and hence, under further assumptions, of the mass of the~Milky~Way.

The escape velocity is the maximum velocity that stars can have while still being bound to the Galaxy. In principle, the single fastest moving bound star places a lower limit on the escape velocity. However, in practice, individual stars might be affected by large measurement uncertainties or they might be outliers (such as escapees). A more robust approach is to fit the velocity distribution as a whole as put forward by \citet[][hereafter LT90]{Leonard1990THESPEED}, who describe the tail of the velocity distribution with a power-law. 

Several works have used the LT90 method in the past. For example, \cite{Smith2007} and \cite{Piffl2014TheWay}, hereafter S07 and P14, estimated the escape velocity locally to lie in the range of $[500-600]$ km/s, using only radial velocity information from RAVE \citep{Steinmetz2006THERELEASE}. The analysis of \citet[][hereafter W17]{Williams2017} supports these values and these authors also show that the escape velocity drops to $\sim 300$ km/s at a distance of 50 kpc. The advent of full phase-space information with {\it Gaia} DR2 has not led to a reduction in the estimated range for the escape velocity in the solar neighbourhood: it is still $[500-640]$ km/s \citep[][hereafter M18 and D19]{Monari2018TheWay, Deason2019TheSpeed}, a result that can largely be attributed to the different underlying assumptions used by the authors.

In this paper, we will use a sample of halo stars with only tangential velocities from {\it Gaia} DR2 to infer the escape velocity applying also the LT90 method. This sample comprises orders of magnitude more halo stars than any other sample used before. Samples making use of only tangential velocities have not been popular for this kind of studies in the past because of the large uncertainties in the velocities, particularly induced by the distance uncertainties. Even more dramatic was the lack of 
(accurate) proper motion measurements for large numbers of stars. However, {\it Gaia} DR2, containing about $\sim 200\times$ more stars with proper motions than radial velocities, makes this kind of study feasible now. We proceed in this work as follows. We describe the data used and its properties in Sec.~\ref{sec:data} and the methods used in Sec.~\ref{sec:method}. In Sec.~\ref{sec:Mockdata} we test the method for determining the escape velocity using mock data and cosmological simulations. In Sec.~\ref{sec:SN} and Sec.~\ref{sec:DR} we present our results in the solar neighbourhood and as a function of galactocentric distance, respectively. In Sec.~\ref{sec:discussion} we use the local escape velocity to derive an estimate of the mass of the Milky Way's dark halo and to identify likely unbound stars. In Sec.~\ref{sec:conclusions} we present our conclusions.

\section{Data}\label{sec:data}
 
The determination of the escape velocity is contingent upon having a
sample of halo stars with high-quality measurements and large velocity
amplitudes. Most of the data used in this work is provided by the {\it Gaia}
mission \citep{GaiaCollaboration2016TheMission,
GaiaCollaboration2018brown}. We will mainly use the sample of halo
stars selected and analysed in \citet[][KH21
hereafter]{Koppelman2020TheCatalogue}. This sample comprises
$\sim 10^{7}$ Main Sequence (MS) halo stars and we refer to it as the reduced proper motion or the 5D sample hereafter. Additionally, we
will also make use of a set of nearby halo stars with full phase-space
information.

\subsection{Velocity information}\label{sec:methvel}

To transform the observed motions (proper motions and radial velocities when available) into space velocities we proceed as follows. We compute the tangential velocity of a star by combining the proper motion and its distance as
\begin{equation}\label{eq:vlvb}
    v_{j} = 4.74057~{\rm km/s}~
    \bigg(\frac{\mu_{j}}{\rm mas/yr}\bigg)~
    \bigg(\frac{d}{\rm kpc}\bigg),
\end{equation}
where $j=(\ell,b)$. These velocities are then corrected for the solar motion using the values for the motion of the Sun with respect to the local standard of rest (LSR) given by \cite{Schonrich2010} and the motion of the LSR given by \citet{Mcmillan2017}; they are $(U_\odot,V_\odot,W_\odot) = (11.1,12.24,7.25)~{\rm km/s}$ and $v_{\rm LSR} = 232.8~{\rm km/s}$ respectively. The transformations to correct the tangential velocities are
\begin{equation}
    v_{j}^\ast = v_{j} + v_{j,\odot},
\end{equation}\label{eq:vlbsolcor}
where $v_{\ell,\odot}$ and $v_{b,\odot}$ are defined as
\begin{subequations}
\begin{equation}
    v_{\ell,\odot} = 
    -U_\odot \sin{\ell} + 
    (V_\odot+v_{\rm LSR})\cos{\ell},
\end{equation}
\begin{equation}
    v_{b,\odot} = 
    W_\odot\cos{b} - 
    \sin{b}\cdot(U_\odot\cos{\ell} + 
    (V_\odot+v_{\rm LSR})\sin{\ell}).
\end{equation}\label{eq:vlvbsun}
\end{subequations}
Finally, the tangential velocity in the Galactic frame of rest as observed from the Sun is calculated as
\begin{equation}
    v_t = \sqrt{ (v_\ell + v_{\ell\odot})^2 + (v_b + v_{b\odot})^2 }.
\end{equation}
Similarly the line-of-sight velocity can be corrected for the solar reflex motion using $v_{\rm los}^\ast =  v_{\rm los} + v_{{\rm los},\odot}$, where
\begin{equation}
    v_{{\rm los},\odot} = 
    {W_\odot}\sin{b} +
    \cos{b}\cdot(U_\odot\cos{\ell} + 
    (V_\odot+v_{\rm LSR})\sin{\ell}).
\end{equation}
To derive space velocities we use the following expressions:
\begin{subequations}\label{eq:vxyz}
\begin{equation}
    v_x = v_{\rm los}^\ast\,\cos{\ell}\cos{b} -v_\ell^\ast\sin{\ell} - v_b^\ast \cos{\ell}\sin{b},
\end{equation}    
\begin{equation}
    v_y =  v_{\rm los}^\ast\,{\sin\ell} \cos{b} + v_\ell^\ast\cos{\ell} - v_b^\ast \sin{\ell}\sin{b},
\end{equation}
\begin{equation}
    v_z = v_{\rm los}^\ast\,\sin{b} + v_b^\ast\cos{b}.
\end{equation}
\end{subequations}
To transform the coordinates to a galactocentric frame we place the
Sun at $X = -8.2$ kpc \citep{Mcmillan2017}. We use this value for the
distance to the Galactic centre because it is consistent with the
\cite{Mcmillan2017} potential that we will employ later, and the same
is true for the LSR velocity. We note however that the
\citeauthor{Mcmillan2017} values agree well with the more recent
determination of the distance to the Galactic Centre by the 
\citet{GRAVITYCollaboration2018DetectionHole} and circular velocity at
the position of the Sun by \citet{Eilers2019TheKpc}.

To isolate a halo sample using the {\it Gaia} DR2 data, we consider stars with velocity vectors that deviate more than $250$~km/s from the velocity vector of the LSR (i.e. the velocity vector of a typical disc star), namely $|\vec{v}- \vec{v}_{\rm LSR}| \geq 250~{\rm km/s}$. This type of selection is known as a `Toomre' selection.

When no line-of-sight velocity information is available, we use Eq.~\eqref{eq:vxyz} setting $v_{\rm los}$ to zero. In that case, we refer to the velocity vector as $(\tilde{v}_x,\tilde{v}_y,\tilde{v}_z)$ to stress that these are not the true Cartesian velocities. For this set of stars, which constitute the majority of our sample, we use an adapted Toomre selection to isolate a halo sample, namely $|\vec{\tilde{v}}- \vec{v}_{\rm LSR}| \geq 250~{\rm km/s}$. 


\begin{figure}
    \centering
    \includegraphics[width=\hsize]{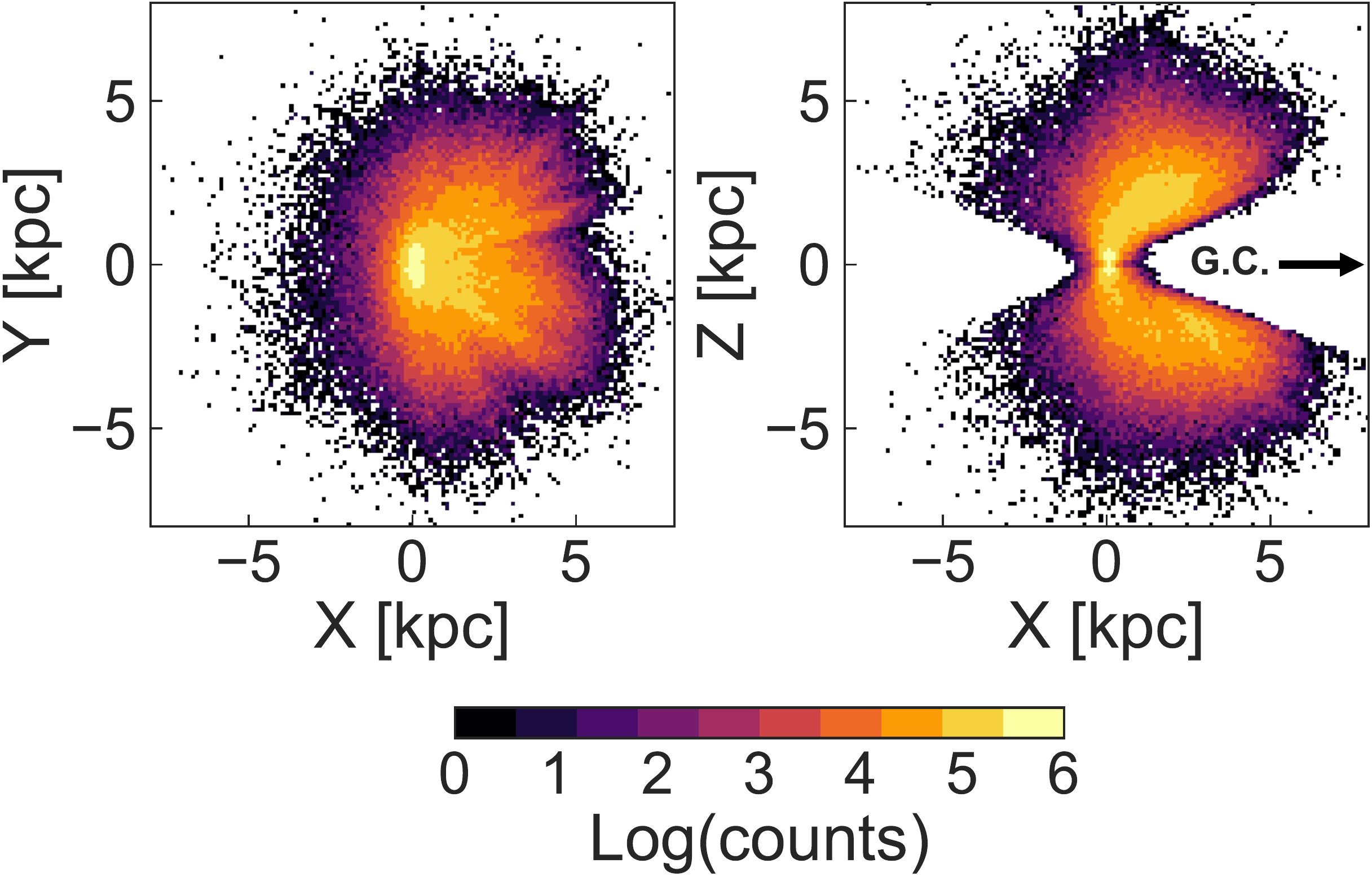}
    \caption{Spatial distribution of the RPM sample used in this work, in heliocentric coordinates and for stars with $v_t > 250$~km/s. The Galactic Centre is located at $X=8.2$~kpc as indicated. The concentration of stars near the origin is caused by a small subset of stars with very accurate trigonometric parallaxes.}
    \label{fig:dataplot1}
\end{figure}

\subsection{Sample with full phase-space information}
\label{sec:6D_sample}
In the solar neighbourhood, we will use a sample of stars with full phase-space information from {\it Gaia} known as the 6D or the radial velocity spectrometer (RVS) sample \citep{Katz2019GaiaVelocities}. We extend this dataset by adding sources with radial velocities observed by APOGEE \citep{Wilson2010TheSpectrograph, Abolfathi2018TheExperiment}, LAMOST \citep{Cui2012}, and RAVE \citep{Kunder2017}, see Sec.~2 of \cite{Koppelman2019a} for a full description of this catalogue. The cross-matches with APOGEE and RAVE have been obtained from the {\it Gaia} archive \citep{Marrese2018}.

\begin{figure*}
    \centering
    \includegraphics[width=\hsize,trim={0 0 4cm 0},clip]{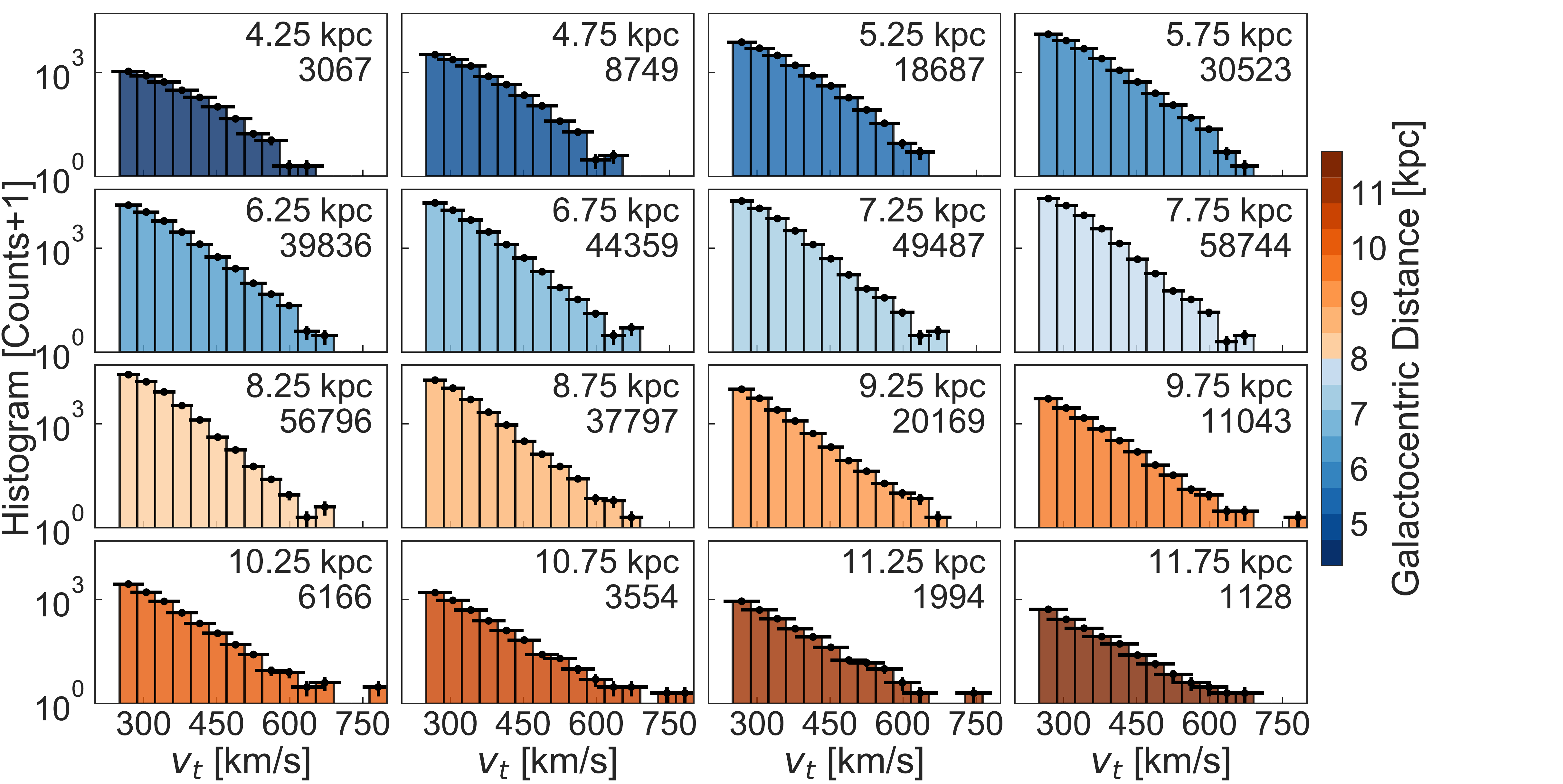}
    \caption{Tail of the tangential velocity distributions for different galactocentric distances. The annotations in the panels indicate the central distance and number of stars per bin. The black marks give the uncertainty in the counts and the mean uncertainty in $v_t$ for each bin.}
    \label{fig:DistRangeHists}
\end{figure*}

For this sample and in line with M18 and D19, we use the quality cuts described in \cite{Marchetti2018}, namely
\begin{itemize}[noitemsep,topsep=0pt]
    \item ${\tt astrometric\_gof\_al}<3$,
    \item ${\tt astrometric\_excess\_noise\_sig}\leq2$,
    \item $-0.23 \leq {\tt mean\_varpi\_factor\_al} \leq 0.32$,
    \item ${\tt visibility\_periods\_used}>8$,
    \item ${\tt rv\_nb\_transits}>5$,
\end{itemize}
and also impose the following quality criteria: 
 \begin{itemize}[noitemsep,topsep=0pt]
     \item ${\tt ruwe<1.4}$,
     \item ${\tt parallax\_over\_error}>5$.
 \end{itemize}
               
For the additional spectroscopic data we use the same quality cuts, with exception of the criterion on {\tt rv\_nb\_transits}. Additionally, we use survey-specific quality constraints. For APOGEE we use 
\begin{itemize}[noitemsep,topsep=0pt]
    \item ${\tt SNR}>20$,
    \item ${\tt STARFLAG}==0$,
    \item ${\rm abs}({\tt SYNTHVHELIO\_AVG}-{\tt OBSVHELIO\_AVG})<50$,
    \item ${\tt NVISITS}>2$,
\end{itemize}
for RAVE 
\begin{itemize}[noitemsep,topsep=0pt]
    \item ${\tt eHRV} <10$,
    \item ${\tt Algo\_Conv\_K} !=1$,
    \item ${\tt SNR\_K} >20$,
\end{itemize}
and for LAMOST
\begin{itemize}[noitemsep,topsep=0pt]
    \item ${\tt snri}>20$,
    \item ${\tt snrg}>20$.
\end{itemize}

Several studies have reported that the sources in the RVS sample, and bright sources in general, contain a parallax offset of $\sim 0.05$~mas \cite[see][]{Schonrich2019, Leung2019SimultaneousLearning, Zinn2019ConfirmationField, Chan2019TheStars}. Therefore, we correct the parallaxes in the 6D sample for an offset of $-0.054$~mas as estimated by \cite{Schonrich2019}. Following these authors, we increase the parallax uncertainties by $0.006$~mas to account for the uncertainties in the offset and by $0.043$~mas to account for the RMS in the offset reported by \cite{Lindegren2018}, both of which are added in~quadrature.

Nonetheless, to mitigate the effects of the parallax offset we only consider sources within 2 kpc. As explained earlier we select halo stars as those with $|\vec{v}-\vec{v}_{\rm LSR}| > 250~{\rm km/s}$. Finally, we remove the star with Gaia DR2 {\tt source\_id} 5932173855446728064 since its radial velocity reported in {\it Gaia} DR2 is known to be incorrect \citep{Boubert2019}. The final sample comprises 2067 high-quality stars, of which 495 are from the {\it Gaia} RVS sample, 10 from APOGEE, one from RAVE, and 1561 from LAMOST. 

Since the spectroscopic surveys add a considerable number of stars, mostly from LAMOST, we have checked that they do not bias our results. In fact, these are fully consistent with using only {\it Gaia} RVS sources. The stars from the spectroscopic surveys do not dominate the determination of $v_{\rm esc}$ because they, in general, have larger uncertainties. However, they do help in closing the confidence contours, as we will see in Sec.~\ref{sec:SN}.

\subsection{The reduced proper motion sample}\label{sec:RPM}

For the complete description of the reduced proper motion (RPM) sample we refer the reader to the KH21 paper. Here we will summarise the details that are relevant for this paper. By virtue of the selection method, the RPM sample comprises only MS stars. The relatively linear colour - magnitude relation for these types of stars can be used to calculate photometric distances with typical uncertainties of $7\%$.

In KH21 we have already introduced several quality cuts, here we prune the sample even further. In summary, we:
\begin{enumerate}[noitemsep,topsep=0pt]
    \item Target the most pure set of halo stars: $|\vec{\tilde{v}} - \vec{v}_{\rm LSR}|>250~{\rm km/s}$ (see Sec.~\ref{sec:methvel}).
    \item Select stars with large tangential velocities: $v_t > 250$ km/s.
    \item Isolate stars that are the least affected by extinction, that is we consider only those with $A_G<0.2$.
    \item Select stars in the colour range where the photometric distances have the smallest uncertainty: $0.50 < {G - G_{\rm RP}} < 0.715$. The blue limit here is stricter than in KH21, to remove any possible contamination from the MS turn-off.
    \item Select stars at high latitudes to remove contamination from the disc: $|b|>20$.
\end{enumerate}

Some stars in the RPM sample have less precise photometric distance than 
trigonometric distance (i.e. parallax from {\it Gaia}). Furthermore, some stars may have been excluded because they did not satisfy the last three quality cuts described~above, even though their trigonometric parallaxes are accurate. Therefore we add such stars back to the sample. We also replace the photometric distances with trigonometric distances for stars with ${\tt parallax\_over\_error}>10$, if the latter has a smaller uncertainty than the first, and we only consider stars with parallaxes~$>0.5~{\rm mas}$.

As mentioned above, the trigonometric parallaxes from {\it Gaia} DR2
are known to contain a zero-point offset that has a
complex dependence on other observational parameters (e.g. the colour
and magnitude of the stars). Because most of the stars in the 5D
sample (without radial velocities) are fainter than those in the 6D
sample, we correct their parallaxes with a different offset. Following
\cite{Lindegren2018}, we use a value of $-0.029$ mas for the parallax
offset and increase the parallax uncertainties by $0.043$ mas (the
uncertainties are added in quadrature) to account for variations in the
offset.

Within 1 kpc about $90\%$ of the distances stem from the {\it Gaia} parallaxes and at 2 kpc this percentage drops to $\sim 50\%$. The final, pruned sample comprises $197~449$ sources of which $18~236$ have {\it Gaia} parallaxes.

\subsection{Inspection of the RPM sample}\label{sec:datainspection}

Figure~\ref{fig:dataplot1} shows the spatial distribution of the stars in the sample. The maps are coloured by the logarithm of the counts per bin. The quality cuts described in the previous section affect the spatial distribution of the stars, most notably by removing low-latitude stars. The overdensity at the solar neighbourhood (centre of the figure) is caused by the addition of sources with accurate parallaxes. The median heliocentric distance of the sample is $3.6$ kpc.

In Fig.~\ref{fig:DistRangeHists} we show the tail of the tangential velocity distribution as a function of galactocentric distance by slicing the sample in uniformly spaced overlapping bins, ranging from $4-12$ kpc, of $1$ kpc width, which is larger than the typical uncertainty in the distances. A visual inspection reveals only small variations in the distributions. These clearly resemble a power-law (as anticipated) but with a slight tendency to become more exponential with distance from the Galactic Centre. 

\begin{figure}
    \centering
    \includegraphics[width=0.9\hsize]{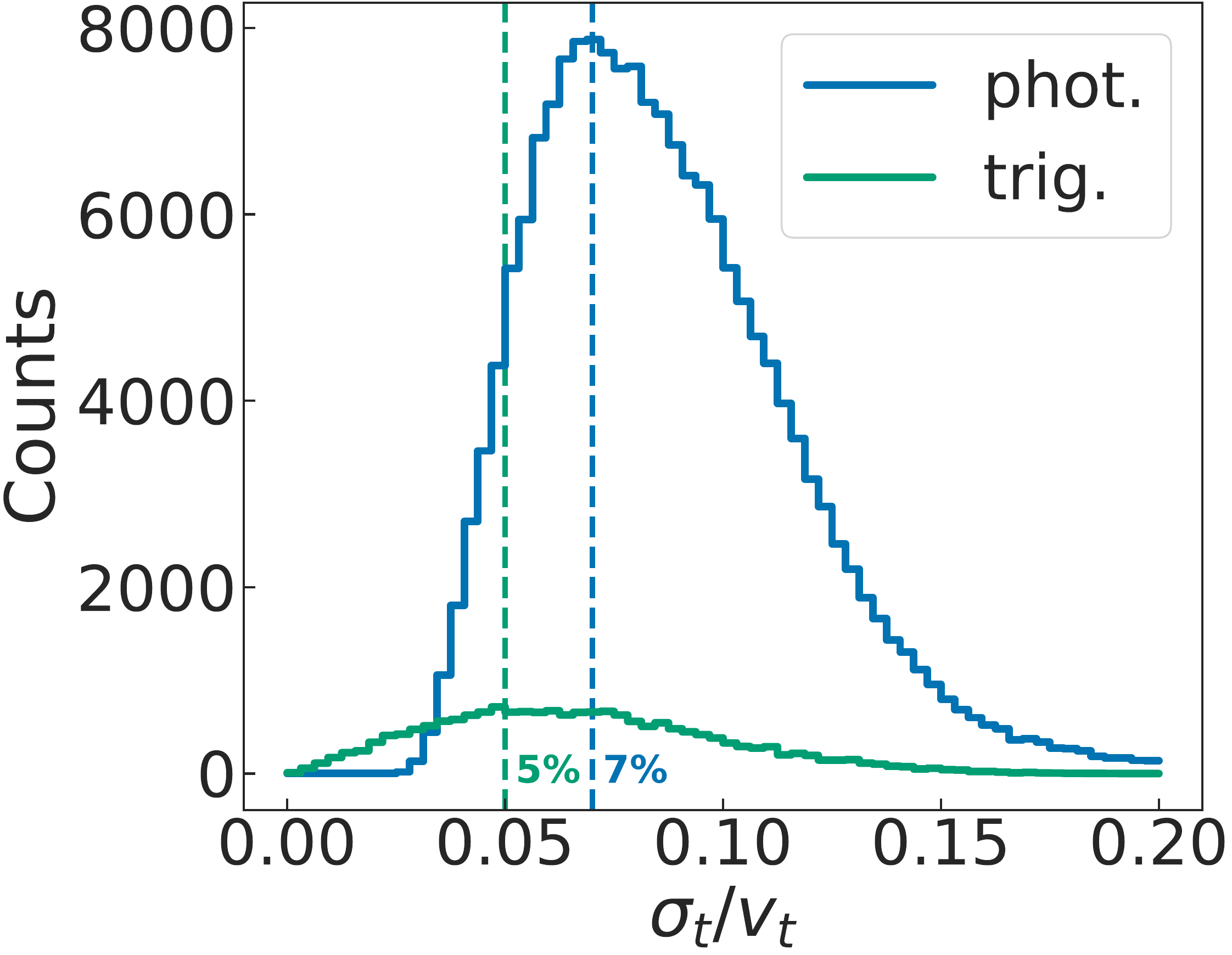} 
    \caption{Distribution of velocity uncertainties for sources in the RPM sample shown separately for sources with photometric (in blue) and trigonometric (in green) distances.}
    \label{fig:velerrors}
\end{figure}

We propagated the uncertainties in the tangential velocities, which we denote as $\sigma_t$, using the standard uncertainty propagation approximation. This approximation uses a Taylor expansion to linearize the coordinate transformations, which is a common practice in the literature. We start the propagation from the measurement uncertainties of Gaia DR2, where the uncertainty of the distance is derived from the parallax uncertainty for the trigonometric sample. For the RPM sample, we derive distance uncertainties separately since this sample contains photometric distances (see Sec.~3.3. of KH21 for more information). Typically, the photometric distance uncertainties are at the $7\%$ level.

The distribution of the relative uncertainties in the tangential velocity $v_t$ is shown in Fig.~\ref{fig:velerrors}, separately for the photometric (in blue) and trigonometric distances (in green). On average, the tangential velocities derived from the trigonometric parallaxes are slightly more accurate than those based on the photometric distances. This is a selection effect since only sources with very accurate parallaxes are included in our sample. When propagating the uncertainties in the velocities, we find that the uncertainty distribution for sources with photometric distances peaks at $7\%$ and has a median of $8\%$. For the trigonometric distances, the distribution in the velocity uncertainties peaks at $\sim 5\%$. The distribution of $\sigma_t$ has a tail towards higher uncertainties because of proper motions uncertainties and there is only is a small dependence with magnitude at the faint end (i.e. for $(G\gtrsim20)$.

\section{Methods}\label{sec:method}

\subsection{Determining $v_{\rm esc}$}\label{sec:meth5D}

As described earlier, we will use here the LT90 method to determine the
escape velocity, denoted hereafter as $v_{\rm esc}$. The motivation of this 
method is that the tail of the velocity distribution can be
described by a power-law, and $v_{\rm esc}$ is the velocity at which
the probability of finding a star goes to zero. Although we follow
closely Sections IIa and IIc from LT90 and adopt their notation, the
formalism we use reveals some differences.

As just stated, the probability of finding a star in a local volume
with a velocity in the range $(v, v+{\rm d} v)$
is described close to the escape velocity as a
power-law
\begin{equation}
    f(v|v_{\rm esc},k) =
  \begin{cases}
    A (v_{\rm esc}-v)^{k}, & \text{$v_{\rm cut} <v < v_{\rm esc}$ ,} \\
                                   0, & \text{$v\ge v_{\rm esc}$,}
  \end{cases}
  \label{eq:distrfunc_full}
\end{equation}
where $k$ is the exponent, $v_{\rm esc}$ is the escape velocity, and $v_{\rm cut}$ is a threshold velocity below which the distribution is not well-represented by a power-law. It is important to set $v_{\rm cut}$ accordingly such that only the tail of the distribution is fit. The normalisation constant is defined as $A = \frac{k+1} {(v_{\rm esc}-v_{\rm cut})^{k+1}}$, which is obtained from $A\,\int_{v_{\rm cut}}^{v_{\rm esc}} f(v|v_{\rm esc},k)\,{\rm d}v = 1$. 

The expression in Eq.~\eqref{eq:distrfunc_full} describes the
distribution of velocities corrected for the solar motion (including
peculiar and LSR), for example at the location of the Sun. Note that if
$f{\rm d}v$ is the probability of finding a star with velocity $v$ in
the range $(v,v+{\rm d}v)$, this implies that there exists some
distribution $g({\bf v})$ such that
$\int_\Omega g({\bf v}){\rm d}{\bf v} = 4 \pi v^2 g(v){\rm d}v =
f(v){\rm d}v$ under the assumption that the velocity distribution is~isotropic.

We now wish to obtain the probability distribution for the tangential velocity (i.e. $f_t(v_t|v_e,k)$). This can be derived from the joint distribution $f_{r,t}(v_r,v_t|v_e,k)$ which gives the probability of finding a star with a given line-of-sight velocity and tangential velocity as $f_{r,t}(v_r,v_t|v_e,k) {\rm d} v_r {\rm d^2} v_t$. By performing a transformation of variables
\begin{equation}
    f_{r,t}(v_r,v_t|v_e,k) = 
    \int g(\vec{v}|v_e,k) \delta(v_r - {{\bf v}\cdot{\bf \hat{n}}})
    \delta(v_t-|{{\bf v} \times{\hat{\bf n}}}|) 
    {\rm d}{\bf v},
\end{equation}
where ${\bf \hat{n}}$ is a unit vector along the line-of-sight. To express the distribution function in terms of only $v_t$ we integrate over the line-of-sight component (and over angle)
\begin{equation}
    f_t(v_t|v_e,k) = 
    \frac{1}{2\pi} \int g(\vec{v}|v_{\rm esc},k) \delta(v_t-|{{\bf v}\times {\bf \hat{n}}}|) 
    {\rm d} {\bf v}.
\end{equation}
The distribution $f_t{\rm d}v_t$ gives the probability of finding a tangential velocity $v_t$ in the range $(v_t,v_t + {\rm d}v_t)$.

Evaluating this integral in spherical coordinates, with ${\bf \hat{n}}$ aligned with the $z$-axis (implicitly assuming the stars are distributed isotropically), we obtain
\begin{equation}
    f_t(v_t|v_e,k) = 
     \iint f(v) \delta(v_t - v\sin{\theta})\,v^2\sin{\theta}\,{\rm d}v{\rm d}\theta,
\end{equation}
which, by changing the order of integration and the substitution of $u = v_t - v\sin{\theta}$, reduces to 
\begin{equation}
    f_t(v_t|v_e,k) = 
    -\int_{v_t}^{v_{\rm esc}} f(v)\,\bigg[v_t^{-2}-v^{-2}\bigg]^{-\frac{1}{2}}\,{\rm d}v.
\end{equation}
The resulting integral for $f(v)$ given by Eq.~\eqref{eq:distrfunc_full} can be solved with {\tt Mathematica} (and depends on the regularised hypergeometric $_2F_1$ function). When evaluating the Taylor series expansion of $v_t \rightarrow v_e$ for the integral, we obtain
\begin{equation}
    f_t(v_t|v_{\rm esc},k_t) \propto
  \begin{cases}
    (v_{\rm esc}-v_t)^{k_t+\frac{1}{2}}, & \text{$v_{\rm cut} < v_t < v_{\rm esc}$,} \\
    0, & \text{$v_t\ge v_{\rm esc}$,}
  \end{cases}
  \label{eq:distrfunc}
\end{equation}
which {the expression found by LT90. It} can be normalised by multiplying with the constant
$A_t = \frac{k_t+1.5} {(v_{\rm esc}-v_{\rm cut}) ^{k_t+1.5}}$, which
is derived from the requirement that
$A_t\,\int_{v_{\rm cut}}^{v_{\rm esc}} f_t(v_t|v_{\rm esc},k_t)\,{\rm
  d}v_t = 1$, and where we have replaced $k$ with $k_t$ for
clarity. The reason for this is that only in the case of
$v_t \rightarrow v_e$ are the two power-law indices of
Eq.~\eqref{eq:distrfunc_full} and Eq.~\eqref{eq:distrfunc} related,
and $k_t = k$. It is thus best to think of $f_t(v_t|v_{\rm esc},k_t)$
in Eq.~\eqref{eq:distrfunc} simply as a power-law description of the
tangential velocity tail, an approximation which is supported by
Fig.~\ref{fig:DistRangeHists}.
We will see in Sec.~\ref{sec:mockplaw} that it is in general not quite
true that $k_t = k$ for the $v_{\rm cut}$ values that are typically
considered in the literature. In what follows, we thus reserve the
notation $k_t$ for the power-law exponent of $f_t$, use $k$ for the exponent
using the distribution from Eq.~\eqref{eq:distrfunc_full} and use
$k_r$ to indicate the exponent for a sample using only line-of-sight
velocities (e.g. when comparing to values in the literature).

So far we have assumed that the velocities ($v$ and $v_t$) are the true velocities. However, in reality we are dealing with `observed' velocities, which are a combination of the true velocity and some unknown uncertainty. In this section we will use $v^\prime_t$ to indicate the observed tangential velocity and reserve $v_t$ for the true value. To account for the uncertainty we smooth the velocities by convolving them with an error distribution $\epsilon(v_t-v^\prime_t, \sigma_t)$, where $\sigma_t$ is the uncertainty in $v^\prime_t$. If
we assume that $v_\ell$ and $v_b$ have Gaussian errors, then the distribution $\epsilon(v_t-v^\prime_t, \sigma_t)$ follows the Beckmann distribution\footnote{The Beckmann distribution is the most general form of the distribution $p(r)$ of parameter $r=\sqrt{x^2+y^2}$, where $x$ and $y$ drawn from a bivariate Gaussian distribution (see \url{https://reference.wolfram.com/language/ref/BeckmannDistribution.html}). The distribution generally can only be expressed in integral form (e.g. Eq. 31 of \citealt{Beckmann1962}) but takes an explicit form for specific cases. For example, when $x$ and $y$ are independent and drawn from a standard normal distribution, $p(r)$ takes the form of the Rice distribution. The more general case of $p(r)$, with $x$ and $y$ are drawn from an uncorrelated bivariate Gaussian distribution, is known as the non-central chi distribution.} (i.e. it is non-Gaussian). However, if evaluated far away from the origin $(v^\prime_t/\sigma_t \gg 0)$, this distribution is well-approximated by a Gaussian. This gives us another reason to choose a sufficiently large $v_{\rm cut}$. Therefore, in what follows we approximate $\epsilon(v_t-v^\prime_t, \sigma_t)$ by a Gaussian~$f_G(v_t-v^\prime_t, \sigma_t)$. The convolution of the power-law from  Eq.~\eqref{eq:distrfunc} and the Gaussian is given by
\begin{equation}
    C(v^\prime_t,\sigma_t,\boldsymbol{\theta}) =  \int_{v_{\rm cut}}^{v_{\rm esc}} f_t(v_t|v_{\rm esc},k_t)f_G(v_t-v^\prime_t,\sigma_t) {\rm d}v_t,
    \label{eq:conv}
\end{equation}
where $\boldsymbol{\theta} = (v_{\rm esc},k_t, v_{\rm cut})$.

We note that we have taken as the integration lower boundary $v_{\rm cut}$ and not zero as in Eq.~(17) of LT90. Since the velocity distribution below $v_{\rm cut}$ is not well-described by a power-law, but by a different distribution function $f^\dagger(v_t)$, the convolution over the range $0 < v_t < v_{\rm cut}$, would take the form 
\begin{equation}
    C^\dagger(v^\prime_t,\sigma, v_{\rm cut}) =  \int_{0}^{v_{\rm cut}}~f^\dagger_t(v_t)\epsilon(v_t-v^\prime_t,\sigma) {\rm d}v_t.
    \label{eq:fullconv}
\end{equation}
which does not depend on $v_{\rm esc}$ nor on $k$. As we will see below, we may thus
ignore this part of the velocity distribution. This also means that we also ignore stars that have an apparent $v^\prime_t$ below the cut, but with a finite probability of having a true $v_t$ above it. We will see in Sec.~\ref{sec:Mockdata} that these assumptions do not affect the method's ability to infer $v_{\rm esc}$. 

By normalising Eq.~\eqref{eq:conv} we find $P(v^\prime_t,\sigma_t|\boldsymbol{\theta})$, the probability of finding a star with $v^\prime_t$ in the range ($v^\prime_t,v^\prime_t+{\rm d}v^\prime_t)$ 
\begin{equation}
    P(v^\prime_t,\sigma_t|\boldsymbol{\theta}) = 
    \frac{C(v^\prime_t,\sigma_t,\boldsymbol{\theta})}
    {\int_{0}^{\infty} C(v^\prime_t,\sigma_t,\boldsymbol{\theta}) {\rm d}v^\prime_t}.
\end{equation}
By definition, because both $f_t$ and $f_G$ are normalised, the integral in the denominator is unity, as the area under a convolution is \mbox{$\int (f\circledast g){\rm d}t = [\int f(u){\rm d}u][\int g(t){\rm d}t] = 1$}. The resulting likelihood function is given by
\begin{equation}
    \mathcal{L} = \prod_{i=1}^{n} P({v^\prime_t}^i,\sigma_t^i|\boldsymbol{\theta}).
\end{equation}

The probability distribution of the model parameters $v_{\rm esc}$ and $k_t$ is found by using Bayes' theorem
\begin{equation}
P(\boldsymbol{\theta}|\Sigma_i^n v_t^i, \sigma_t^i) \propto  P(v_{\rm esc})P(k_t) \prod_{i=1}^{n} P({v^\prime_t}^i,\sigma_t^i|\boldsymbol{\theta}),\label{eq:Prob_ve_k}
\end{equation}
where $P(v_{\rm esc})$ and $P(k_t)$ are priors for $v_{\rm esc}$ and $k_t$. For numerical reasons the logarithm of the probability is evaluated, which also allows us to ignore the normalisation since that is constant and independent of the model parameters. 

The procedure that is outlined above implicitly makes the following assumptions:
\begin{enumerate}[noitemsep,topsep=0pt]
    \item The tail of the velocity distribution is populated up to the escape velocity.
    \item The tail of the velocity distribution is smooth. 
    \item There are no unbound stars in our sample. 
    \item And there is no contamination from a rotating (disc-like) population, which would break the isotropy on the sky.
\end{enumerate}
Perhaps the most problematic assumption is the first one. There is no guarantee that the velocity distribution locally, or at any other location in the Milky Way, extends up to the escape velocity. Most likely it is truncated at some lower value. As a result, the LT90 method is prone to underestimate the true $v_{\rm esc}$. For example, cosmological simulations show velocity distributions that are truncated at $90\%$ of $v_{\rm esc}$ (e.g. S07). The exact location of the truncation depends on the assembly history of the galaxy and quite possibly also on the resolution of the simulation. We will quantify the truncation of the velocity distribution using mock data in Sec.~\ref{sec:Mockdata}. We stress that, because of this truncation, whatever value we derive for $v_{\rm esc}$ it most likely is a \emph{lower~limit}. 

The second assumption has recently been tested by \cite{Grand2019TheHalo}, who find that clustering in the velocity distribution biases the estimation of $v_{\rm esc}$, and can result both in under and overestimates. Nonetheless, these authors show that the estimated $v_{\rm esc}$ is typically underestimated by $7\%$. To emphasise the importance of this bias: a difference of $7\%$ in the escape velocity results in a $21\%$ bias in the estimated mass.

It seems unlikely that our sample contains many unbound
stars, since hyper-velocity stars are typically young stars ejected from
the Galactic Centre and not old stars in the halo
\citep[e.g.][]{Brown2015HypervelocityStars, Boubert2018}. Furthermore, the velocity distributions shown in Fig.~\ref{fig:DistRangeHists} are relatively smooth, suggesting the
presence of a single population dominated by main sequence halo stars. 
Nonetheless, it would be interesting to follow-up spectroscopically stars near the escape velocity. In Sec.~\ref{sec:unbound} we will revisit
possible outliers in the solar neighbourhood. 

\subsection{Adopting a prior on $v_{\rm esc}$ and $k_t$}\label{sec:priors}
In line with the literature, we assume a simple prior on $v_{\rm esc}$
of the form $P(v_{\rm esc}) \propto 1/v_{\rm esc}$. For $k$ (we will use the notation $k$ here,
understanding that it only compares to $k_t$ and $k_r$ in the limiting case)
there is some debate in the literature on
what to assume, and since $v_{\rm esc}$ and $k$ are highly degenerate (see next section),
the prior assumed might bias the resulting $v_{\rm esc}$. For example,
the M18 and D19 estimates of $v_{\rm esc}$ differ by
$\sim 50~{\rm km/s}$ mainly because of the different ranges considered for $k$. Attempting to estimate $v_{\rm esc}$ and $k$
simultaneously is only possible with a large sample with very small
uncertainties. For example, LT90 estimated that a sample of $>200$ stars with high-quality radial-velocities above $v_{\rm cut}$ is necessary to estimate both values simultaneously.

LT90 argue that $k$-values should be in the range $[0.5-2.5]$, because this brackets $k=1.5$, which is the value expected for a system that has undergone violent relaxation \citep{Aguilar1986THEGALAXIES, Jaffe1987TheGalaxies, Tremaine1987Summary}. S07 have compared stellar halos in cosmological simulations of Milky Way-like galaxies and found a range of $[2.7 - 4.7]$ to be more appropriate. P14 building on more recent such simulations reduced this range to $[2.3 - 3.7]$, which is also the range used by M18. D19
updated the criteria for finding Milky Way analogues based on recent discoveries regarding the merger history of the Milky Way \citep{Belokurov2018Co-formationHalo, Helmi2018}. When using cosmological simulations, the range $[1.0 - 2.5]$ was found to be more favoured. Using a sample of BHB stars, K-giants, and MSTO stars from SDSS with only line-of-sight velocities, W17 determine both $k$ and $v_{\rm esc}$ simultaneously. They report a value for $k_r$ of $4\pm1$ for the local stars. 

The above paragraph shows that no consensus has been reached on the value of $k$ for the Milky Way. To complicate matters, the ranges mentioned above were determined for the stellar halo at the position of the Sun (in the simulations). It is not clear whether $k$ remains constant as a function of distance to the Galactic Centre. In this work, we will rely mainly on the estimate of $k$ at the location of the Sun. This is where our sample contains many stars with reliable parallax information and which are approximately isotropically distributed on the sky. For this local sample of stars, we calculate the marginalised posterior distribution for $k_t$. We will apply this posterior as a prior to other distance bins in which we estimate $v_{\rm esc}$. In doing so, we assume that $k_t$ does not vary (much) over the distance range that we probe, which is also justified by the analysis we carry out in Sec.~\ref{sec:Aurigaia}.

\section{Validating the method}\label{sec:Mockdata}
Before applying the method to the data we will attempt to establish the accuracy of the method, the sample size required to estimate both $v_{\rm esc}$ and $k_t$ at the
same time, and the effect of the cut-off parameter $v_{\rm cut}$. We first look at mock data and then apply the method on cosmological simulations.

\begin{figure}
    \centering
    \includegraphics[width=\hsize]{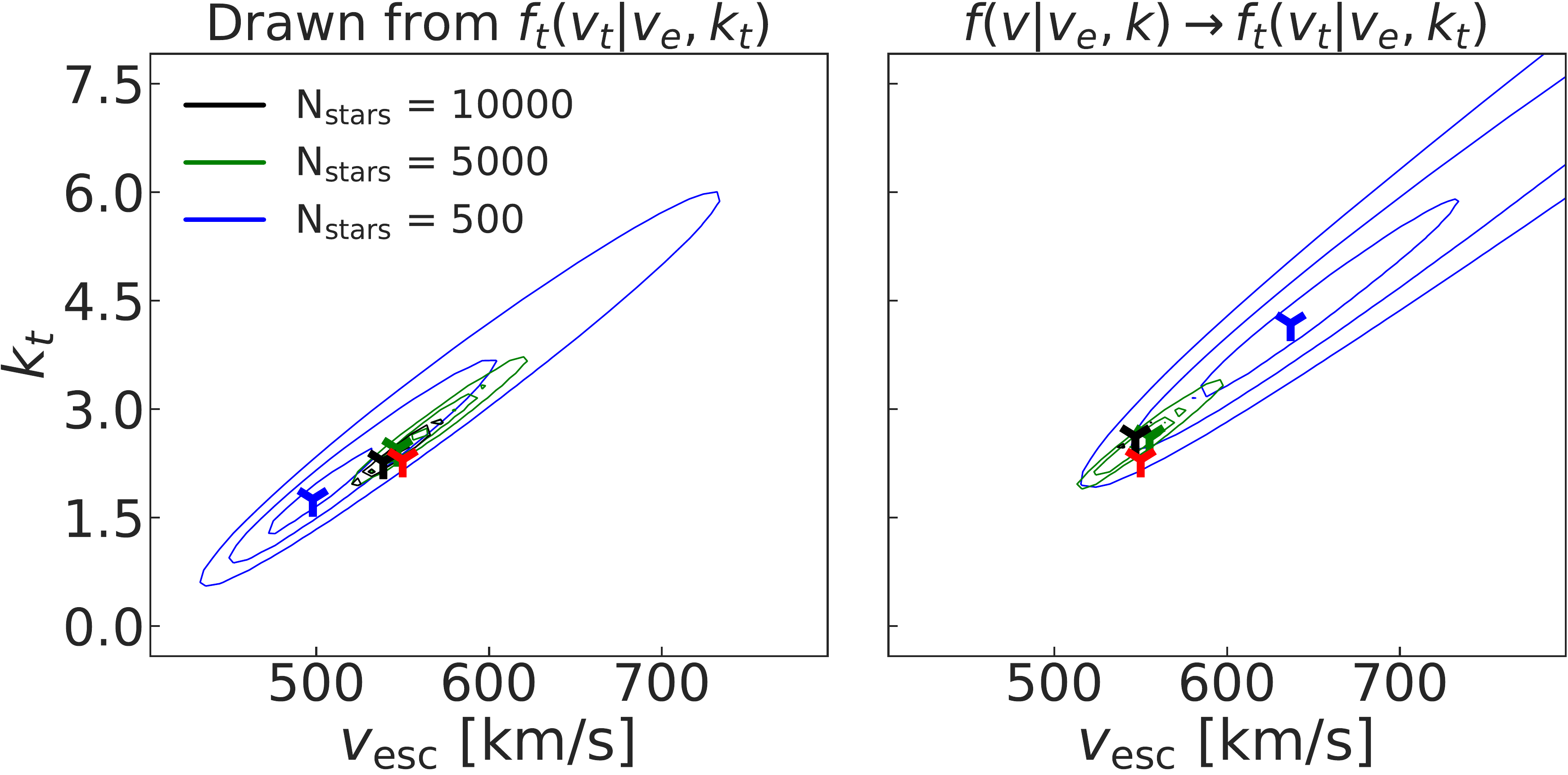}
    \caption{Probability distributions of $v_{\rm esc}$ and $k_t$
      derived using tangential velocities for a mock data sample drawn
      from a power-law distribution in the velocity modulus, and convolved with realistic
      uncertainties. The left panel shows the results based on mock data drawn directly from Eq.~\eqref{eq:distrfunc} whereas on the right we draw the data from Eq.~\eqref{eq:distrfunc_full} and then transform the velocities to tangential velocities. The closed contours mark the probability levels where the probability has dropped to 61\%, 14\%, and 1\% of the maximum a posteriori value (they correspond to the $1$, $2$, and $3\mbox{-}\sigma$ levels if the distribution were Gaussian). The coloured markers indicate the highest probability parameter combinations, with the red marker showing the input parameters.}
    \label{fig:mock-powlaw}
\end{figure}

\subsection{Tests with mock data}\label{sec:mockplaw}

The mock data are drawn from an idealised power-law distribution. We
sample velocities according to the distribution given by Eq.~\eqref{eq:distrfunc} assuming $v_{\rm esc} = 550~{\rm km/s}$
and $k_t = 2.3$.
We convolve the resulting distribution with realistic uncertainties drawn from the
distribution of uncertainties (i.e. that shown in Fig.~\ref{fig:velerrors}), for the
photometric distances sample. 

The left panel of Fig.~\ref{fig:mock-powlaw} shows the results of applying the
formalism described in Sec.~\ref{sec:method} to this dataset for three
different sample sizes (see annotation) above
$v_{\rm cut} = 250~{\rm km/s}$. The contours mark the $1,2,3\mbox{-}\sigma$ levels estimated by the level where the probability
has dropped to $61\%$, $14\%$, and $1\%$ of the probability of the most likely parameter combination.
The true $v_{\rm esc}$ and $k_t$ of the parent distribution are marked
with a red marker. Decreasing the
number of stars (from $10~000$ to $500$) results in higher
uncertainties in the estimates of the $v_{\rm esc}$ and $k_t$ parameters. A
sample with $\sim 10^4$ stars is sufficiently large to determine both
$v_{\rm esc}$ and $k_t$ at the same time, given the amplitude of the
velocity uncertainties.

\begin{figure}
    \centering
    \includegraphics[width=\hsize]{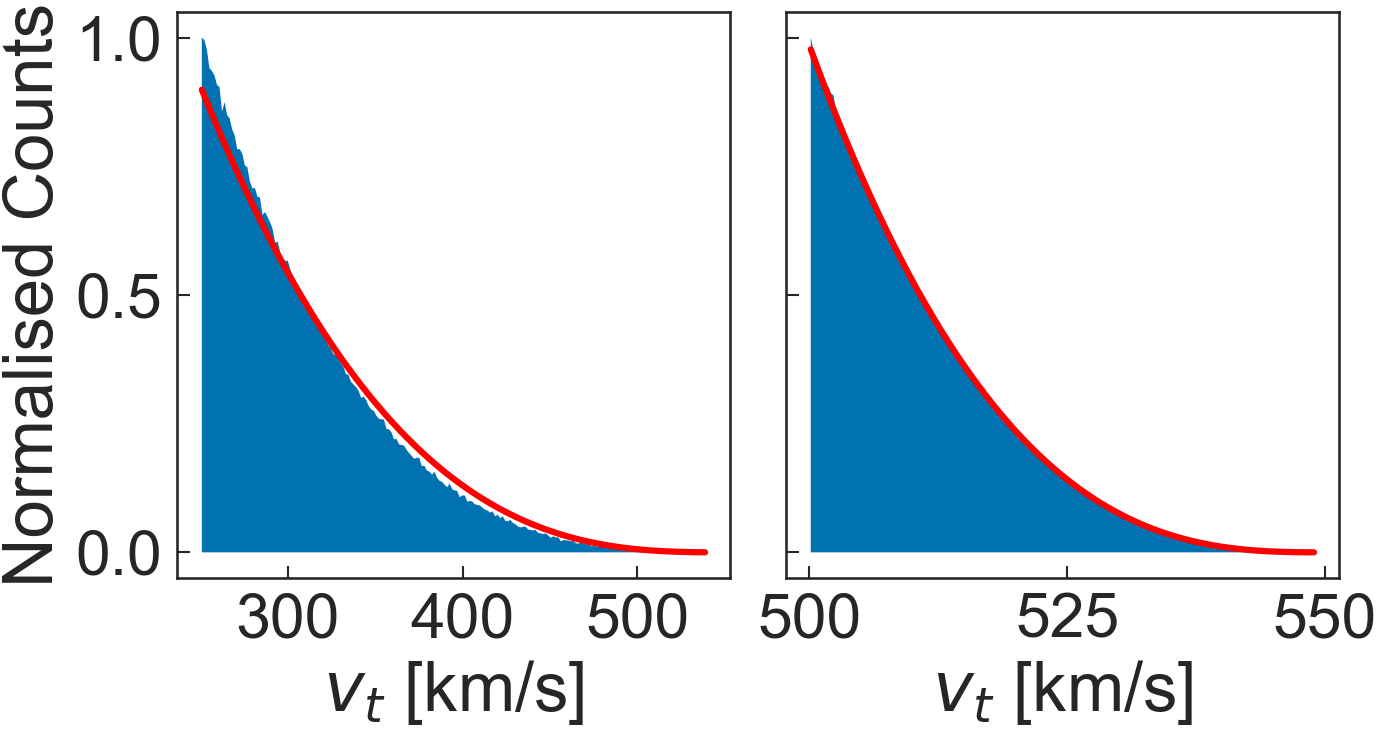}
    \caption{Mock tangential velocity distributions drawn using
      Eq.~\eqref{eq:distrfunc_full}, for two cut-off velocities
      (250, 500) km/s (left and right panels, respectively). The
      red line indicates the distribution that is expected when
      $v_t \rightarrow v_{\rm esc}$ (i.e. Eq.~\ref{eq:distrfunc}).}
    \label{fig:mock-vcut}
\end{figure}

We also tested a procedure drawing the mock distribution of $v_t$ starting from a parent distribution of full 3D velocities (e.g. starting from Eq.~\ref{eq:distrfunc_full} rather than from Eq.~\ref{eq:distrfunc}). The 3D velocities were then transformed to $v_t$ velocities by artificially setting one component to zero, assuming that the velocities are distributed isotropically. Arguably, the resulting distribution of mock $v_t$ more closely describes the observed distribution of $v_t$ than the one drawn directly from Eq.~\eqref{eq:distrfunc}. The right panel of Fig.~\ref{fig:mock-powlaw} shows that the resulting values for $v_{\rm esc}$ are close to the input value. However, the values for $k_t$ are slightly overestimated. This overestimate arises from the difference between $k$ and $k_t$ for low $v_{\rm cut}$ and was already anticipated in Sec.~\ref{sec:meth5D}. 

To emphasise this behaviour, we show in Fig.~\ref{fig:mock-vcut} the behaviour of the artificial $v_t$ distribution (drawn without uncertainties) compared to, in red, the expected distribution of $k_t \rightarrow k$ if $v_t \rightarrow v_{\rm esc}$ (i.e. Eq.~\ref{eq:distrfunc}). The two distributions are only equivalent for cut-off velocities $v_{\rm cut}$ very close to the escape velocity.

Although this does not invalidate
our approach at all since $v_{\rm esc}$ is robustly determined without
any biases, we nonetheless have to be cautious when comparing the
value of $k_t$ obtained using tangential velocities. Similar considerations are in order when applying the LT90 method to a sample
of line-of-sight velocities only. 

\subsection{Tests on Aurigaia Milky Way-like halos}\label{sec:Aurigaia}

We now test the method on two halos from the Aurigaia suite of mock
{\it Gaia} catalogues \citep{Grand2019TheHalo}. We explore here
whether the tail of the velocity distribution is well described by a
power-law, the effect of velocity clumps, the behaviour of $k_t$ as a
function of distance, and the power of the method given the typical
uncertainties in the tangential velocity in our sample.

The Aurigaia catalogues have been generated from the Auriga suite of Milky
Way-like galaxies \citep{Grand2017TheTime} -- which is a suite of
high-resolution, zoom-in re-simulations based on galaxies extracted
from the EAGLE simulations \citep{Schaye2014}. The mock catalogues
that we analyse correspond to halos $6$ and $27$, have the bar at $30$
degrees orientation, and were generated with the {\tt SNAPDRAGONS}~code 
\citep{ Hunt2015TheObservations}. We will refer to these simulations
as Au-06 and Au-27. These specific halos are chosen somewhat at
random, although Au-06 is `the closest example to the MW according to
halo spin' according to \cite{Grand2018Aurigaia:Simulations}. The halo
of Au-06 has a similar mass as the Milky Way (i.e.
$M_{200} \approx 10^{12}~{\rm M}_\odot$), whereas that of Au-27 is
slightly more massive: $M_{200} \approx 1.7\cdot10^{12}~{\rm M}_\odot$
\citep{Grand2017TheTime}. Both halos are mildly prograde ($\sim 30-70$
km/s), as measured by the mean rotational velocity of accreted stars with
`heliocentric' distances smaller than 1 kpc.

Because the original Auriga simulations do not have the resolution of
{\it Gaia} DR2 ($\sim 10^9$ stars), the {\tt SNAPDRAGONS} code has
been used to artificially increase the number of objects, whereby
simulated stellar particles are split into multiple `stars'. This
leads to artificial enhancement of the clustering of stars in
phase-space, which can lead to biases in the determination of the
escape velocity. Therefore, here we only use unique stellar particles by filtering
all duplicates using the true {\tt HCoordinates} and {\tt HVelocities}
parameters in the Aurigaia catalogue.

\subsubsection{The high-velocity tail in Aurigaia}

\begin{figure}
    \centering
    \includegraphics[width=0.8\hsize]{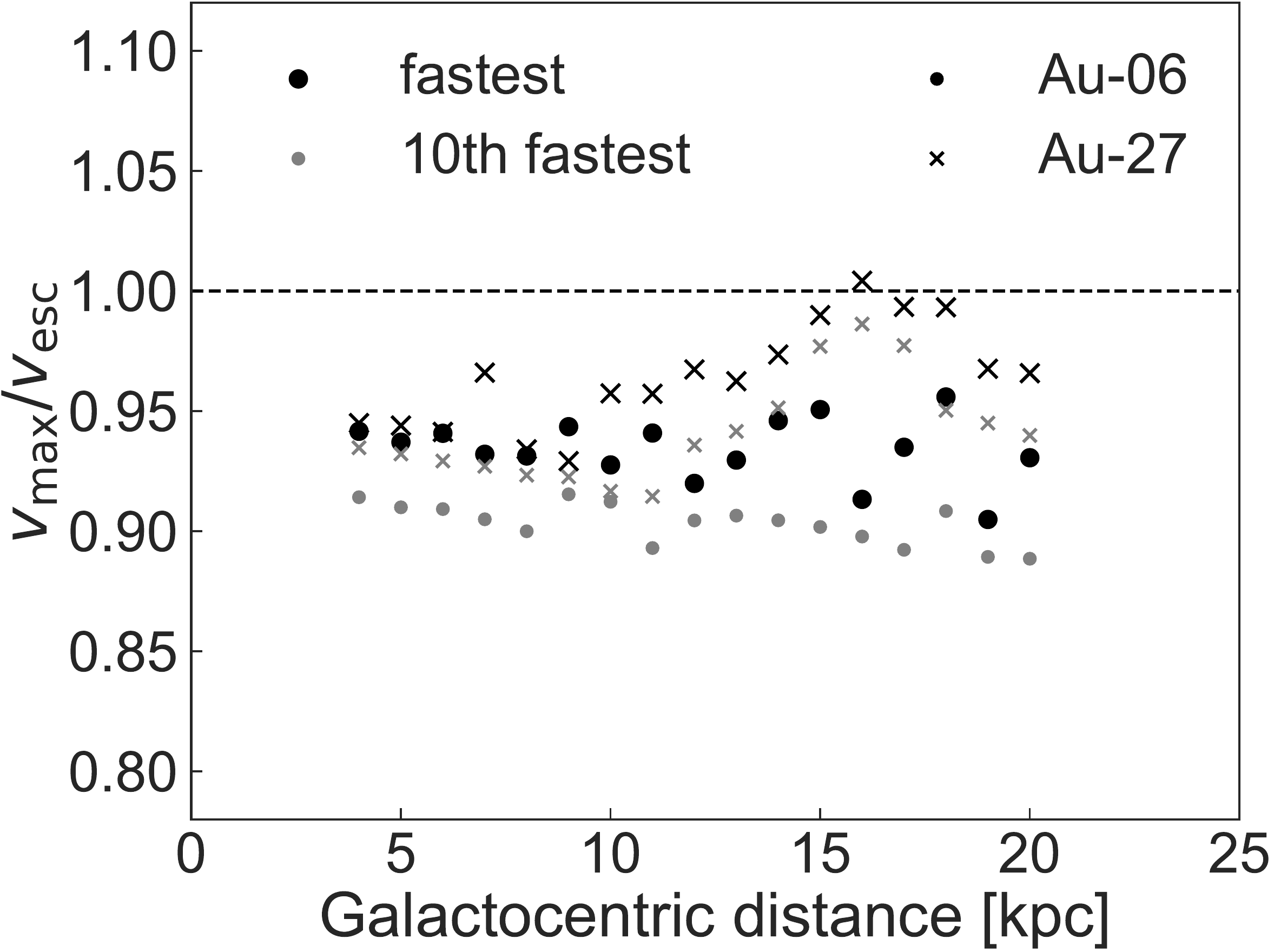}
    \caption{Truncation of the velocities in the Aurigaia halos as a function of galactocentric distance. The velocity distribution in the halos is truncated at $\sim 95\%$, except for a few bins in the outer regions of Au-27. The black markers show the stars that are the closest to the escape velocity. To indicate how densely populated the high-velocity tail also the 10th star fastest star is shown (grey markers).}
    \label{fig:Aurigaia-vtrunc}
\end{figure}

Figure~\ref{fig:Aurigaia-vtrunc} shows the velocity of the fastest moving stars relative to the escape velocity and as a function of distance. We see that typically, the fastest star moves at $90-100\%$ of the true $v_{\rm esc}$ throughout the range of galactocentric distances probed. 
The escape velocity has been calculated here as the velocity needed to reach $r\to\infty$ for the potential given in the Aurigaia catalogue (parameter: {\tt GravPotential}), and using
\begin{equation}
    v_{\rm esc}(r\to\infty) \equiv \sqrt{2|\Phi(r)-\Phi(\infty)|} = \sqrt{2|\Phi(r)|}.
    \label{eq:vescinfty}
\end{equation}
In Fig.~\ref{fig:Aurigaia-vtrunc} the black markers correspond to the fastest star while the grey markers indicate the location of the 10th fastest star and provides an idea of the steepness of the velocity tail.

Interestingly, for the halo of Au-27 the velocity distribution is truncated close to the escape velocity around a galactocentric radius of $\sim 15~{\rm kpc}$. 
In the inner regions, particularly for Au-6 but also to some extent for Au-27, the difference between the fastest and the 10th fastest moving star shows typically less scatter, indicating that there are many stars near the truncation of the velocity distribution.

\subsubsection{Determination of the escape velocity in Aurigaia}

We follow a
similar procedure as for the data to select stars with large
tangential velocities from the Aurigaia halos. Firstly, the tangential velocities are
convolved with uncertainties drawn from the `observed' uncertainty distribution
shown in Fig.~\ref{fig:velerrors} (and as in Sec.~\ref{sec:mockplaw}). We
then select stars that have $v_t>200~{\rm km/s}$. Next, we
artificially set the line-of-sight velocities to zero and select stars
with $|\vec{\tilde{v}} - \vec{v}_{\rm LSR}| > 250~{\rm km/s}$ (as in
Sec.~\ref{sec:RPM}). Although the Aurigaia catalogues do not exactly represent
the Milky Way, these velocity cuts serve to remove the thin disc and
(most of) the thick disc present in the simulations.

\begin{figure}
    \centering
    \includegraphics[width=\hsize]{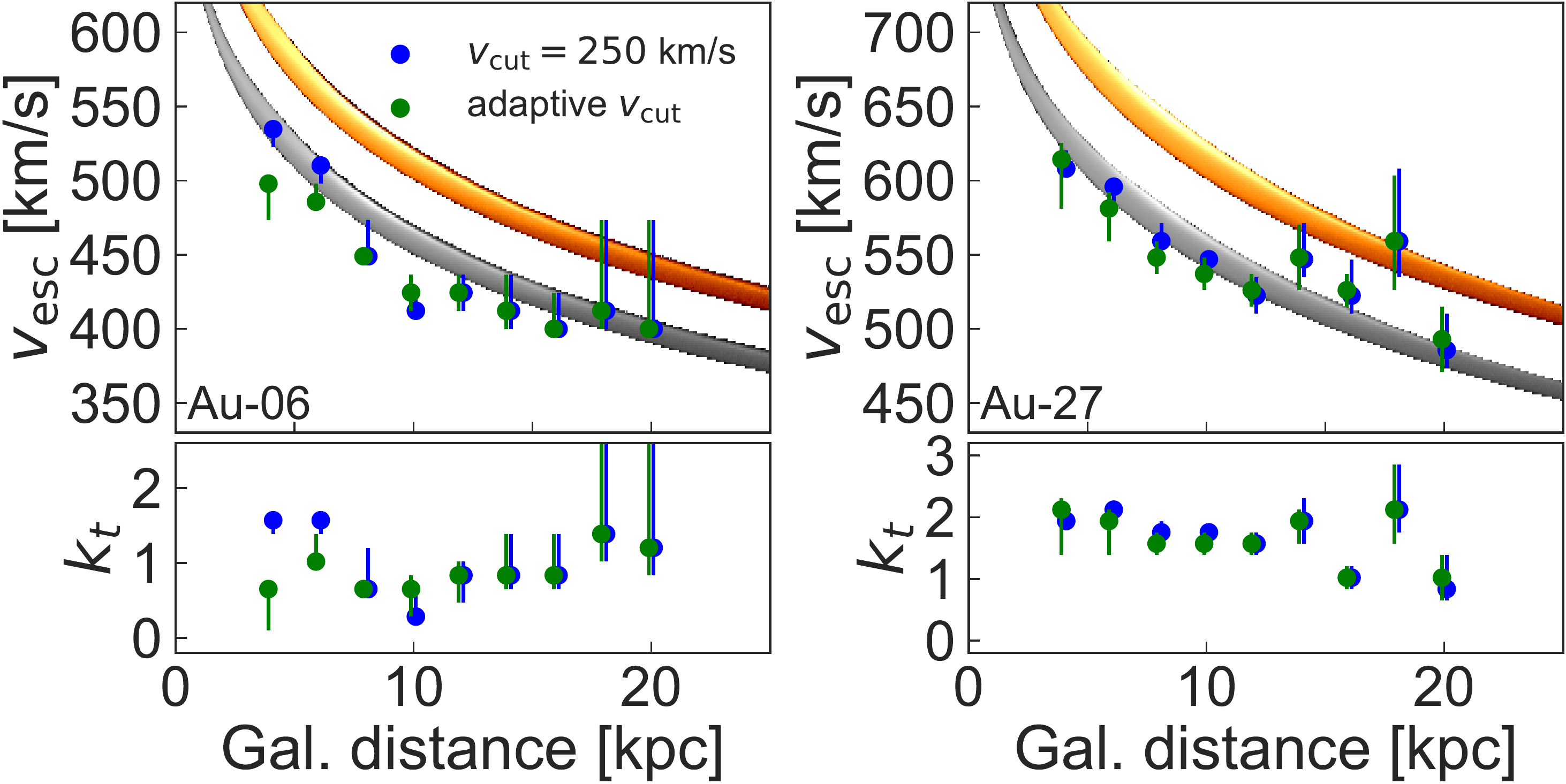}
    \caption{Determination of $v_{\rm esc}$ and $k_t$ as a function of galactocentric distance. The results for both a fixed (blue) and adaptive (green) cut-off velocity are shown. The yellow contour in the background shows the true escape velocity, calculated as $\sqrt{2|{\tt GravPotential}|}$. The grey contour has been shifted downwards by 10\%. The error bars indicate the 3-$\sigma$ confidence levels.}
    \label{fig:Aurigaia-combined}
\end{figure}

We then determine $v_{\rm esc}$ and $k_t$ in concentric shells of 1
kpc in width centred on the galaxy's centre, with radii ranging from
$2 - 21$ kpc. For both Auriga halos the cut-off velocity is set at
$v_{\rm cut} \approx 250$ km/s. This is well below the escape velocity
in all the distance bins we probe. We also test a heuristic procedure
to determine $v_{\rm cut}$ by taking the maximum of 250 km/s and the
velocity of the 10~000th fastest moving star (20~000th for Au-27). For
bins with a large number of stars, this pushes the cut-off to higher values.

We noticed that for the Au-27 halo, the top 20~000 stars works better
to determine $v_{\rm cut}$ than the top 10~000. 
Since this halo is more massive than that of
Au-06, its escape velocity is higher and there
are more stars with extreme velocities. However, because the Au-06 halo is more
similar to the Milky Way, we expect that 10~000 is a realistic number
of stars for the Milky Way.

\begin{figure}
    \centering
    \includegraphics[width=\hsize]{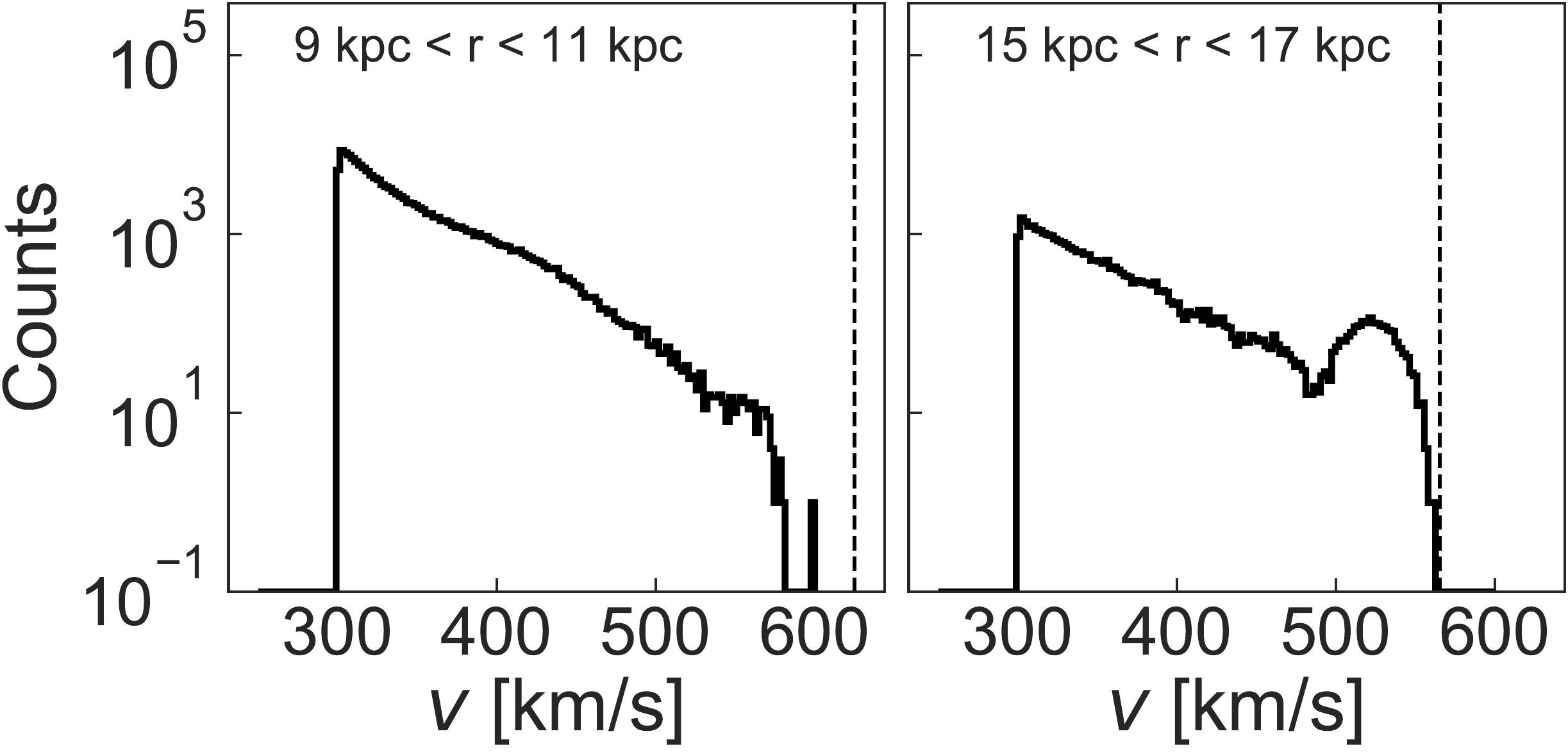}
    \caption{Velocity distribution in Au-27 in two distance ranges. Left: a smooth distribution of stars in the range $9-11$ kpc that is truncated shortly before the escape velocity indicated by the dashed vertical line. Right: a clearly non-smooth velocity distribution for stars in the range $15-17$ kpc that reaches up to the escape velocity. This figure shows the full velocities (and not $v_t$) to emphasise the clumpiness.}
    \label{fig:Au27-bump}
\end{figure}

Figure~\ref{fig:Aurigaia-combined} shows the results of fitting the
tangential velocity tail in the halos of Au-06 (left) and Au-27
(right). In yellow we show the mean $v_{\rm esc}$ that is calculated
from Eq.~\eqref{eq:vescinfty} by using the pre-computed potential
energies of every particle (i.e. the {\tt GravPotential}
parameter). The width of the yellow region indicates that there is a
range of escape velocities at a fixed radius. This range exists
because the potential is not spherically symmetric. Stars
close to the disc experience a stronger potential than those slightly farther away.

The results obtained from a fixed $v_{\rm cut}$ are indicated with
blue markers, while green markers are for the adaptive $v_{\rm
  cut}$. The top panels of Fig.~\ref{fig:Aurigaia-combined} show that
the estimates are systematically too low compared to the expected
$v_{\rm esc}$ for both halos. However they match well with the grey
curve which has been obtained by lowering by 10\% the yellow
region. This is a reflection of a truncation in the tangential
velocity distribution, in that it does not extend all the way to
$v_{\rm esc}$. Note that in both halos, the features in the
$v_{\rm esc}$ curve are matched closely by the velocity features
apparent for the 10th fastest stars shown in
Fig.~\ref{fig:Aurigaia-vtrunc}. An interesting result is that $k_t$
varies only weakly over distance as can be seen from the bottom panels
of Fig.~\ref{fig:Aurigaia-combined}.

The above results mean that the determination of $v_{\rm esc}$ with
the method described in Sec.~\ref{sec:method} is sensitive to the
behaviour of the tail of the velocity distribution. This is
particularly clear for Au-27 which shows a bump in $v_{\rm esc}$ at
$d \gtrsim 15~{\rm kpc}$. In fact there is an excess of stars (a
clump) in the halo of Au-27 that is moving at a velocity close to
$v_{\rm esc}$ as can be seen by comparing the panels of
Fig.~\ref{fig:Au27-bump}, which plot the velocity distributions for the
distance ranges $9-11$ kpc and $15-17$ kpc.

In summary, the analysis of the Aurigaia experiments analysis shows that 
\begin{itemize}[noitemsep,topsep=0pt]
    \item $v_{\rm cut}$ can be determined from the top 10~000 stars.
    \item We may assume that $k_t$ varies only weakly over the distance range probed by the RPM sample.
    \item On average the method underestimates $v_{\rm esc}$ by
      $\sim10\%$. This is slightly more than the 7\% estimated by
      \cite{Grand2019TheHalo}, which might be related to differences
      in the method (e.g. the convolution with an uncertainty distribution and the typically large uncertainties on $v_t$).
    \item By determining $v_{\rm esc}$ over a range of galactocentric distances we can check for local `biases'.
\end{itemize}

\section{Results: solar neighbourhood}\label{sec:SN}

\begin{figure}
    \centering
    \includegraphics[width=\hsize]{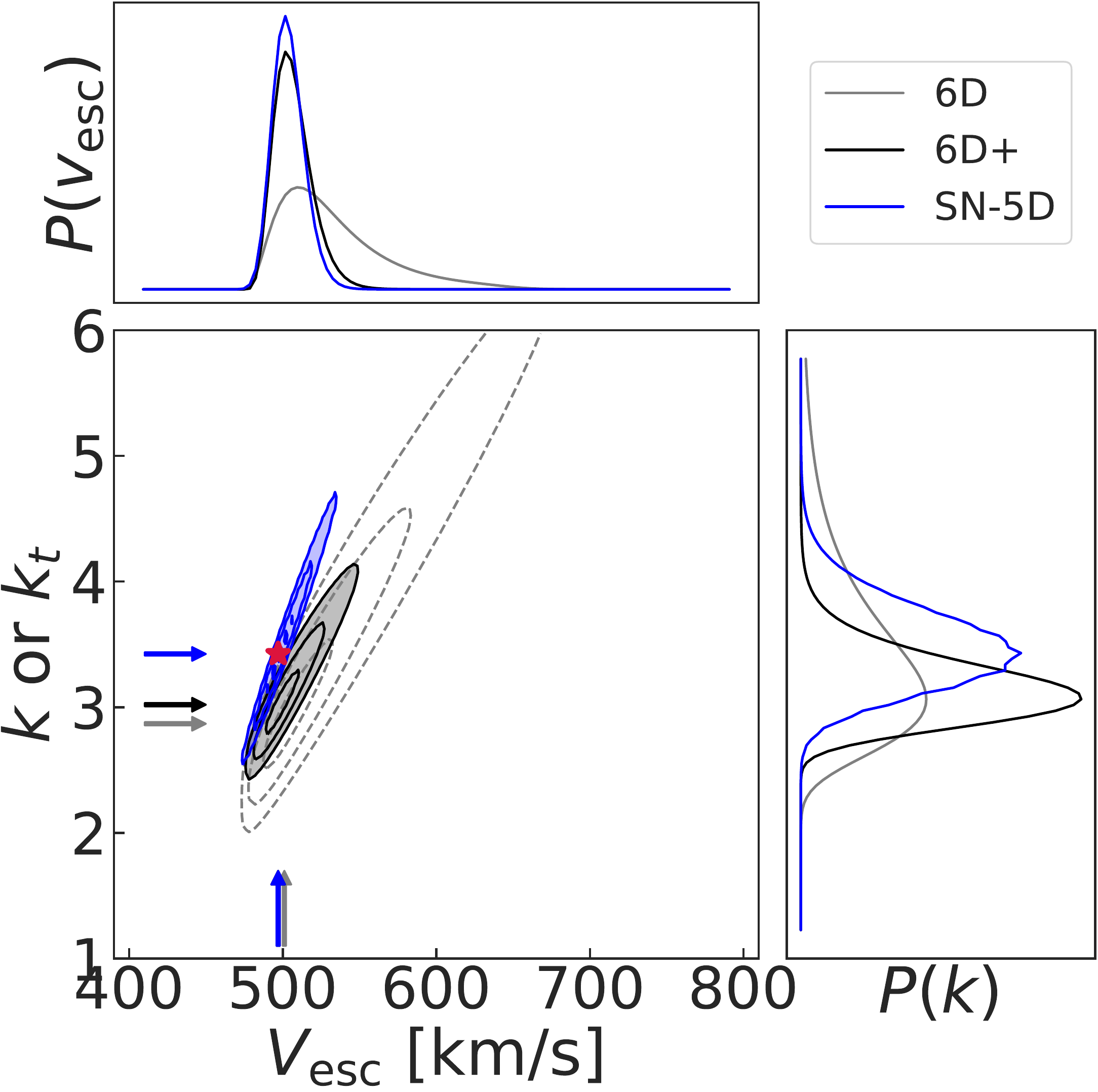}
    \caption{Confidence levels obtained by applying the LT90 method to the 5D and 6D samples in the solar neighbourhood. For each curve the $1$, $2$, and $3\mbox{-}\sigma$ levels are shown and the arrows indicate the maximum probability values. The side panels show the marginalised posterior distributions for $P(v_{\rm esc})$ and $P(k)$. For the 6D sample we show the results for both the augmented dataset and when using {\it Gaia} data only. For the 5D sample, recall that the method determines $k_t$, and this is what is shown on the $y$-axis of the main panel, while the blue curve in the right panel represents $P(k_t)$.}
    \label{fig:SNcombined}
\end{figure}

We determine the escape velocity at the solar position using the two samples of stars described in Sec.~\ref{sec:data}, one with full 6D information and the other with only tangential velocities (5D). We consider only stars with a heliocentric distance of 2~kpc or less. We evaluate the probability (Eq.~\ref{eq:Prob_ve_k}) on a grid of $100\times100$ points ranging from $400~\mathrm{km/s}<v_{\rm esc}<800~\mathrm{km/s}$ and $1<k_t<6$ (both for the 5D and 6D cases). These ranges bracket the values that are presented in the literature. For the 5D sample the cut-off velocity is based on the 10~000th fastest star and set to $317$ km/s, and for the 6D sample it is $250$ km/s. Although the results for the 6D sample are consistent when $v_{\rm cut}$ is set to 317 km/s, in this case the inference on $k$ is less strong. 

The confidence contours for the two samples are presented in
Fig.~\ref{fig:SNcombined}. For the sample with full phase-space
information we plot the results for the {\it Gaia}-only data (6D) and
also including the additional data from ground-based spectroscopic
surveys (6D+). The arrows in the
figure indicate the maximum probability values for each sample. The
contours correspond to estimates of the $1,2,3\mbox{-}\sigma$ levels
(see Sec.~\ref{sec:mockplaw}). The side-panels show
the marginalised distributions $P(v_{\rm esc})$ and $P(k)$ ($P(k_t)$
for the 5D sample). These distributions are the best constrained for
the 5D sample (in blue) because of its large number of stars.

The marginal distributions of $v_{\rm esc}$ agree very well with
each other for all samples. The slight tension in the 5D and 6D curves
(the contours are however consistent within the $2\mbox{-}\sigma$ level),
is driven by the anticipated differences that are the result of using the full
velocity modulus or tangential velocity information only (i.e.
$k \ne k_t$ for $v_{\rm cut}$ far from $v_{\rm esc}$, c.f. the
contours and red marker in Fig.~\ref{fig:mock-powlaw}).

Marginalising over $k_t$, we find a maximum probability value $v_{\rm esc} = 497^{+8}_{-8}~{\rm km/s}$ for the 5D sample, which we stress is most likely biased low compared to the actual $v_{\rm esc}$. The quoted uncertainties correspond to the marginalised $68\%$ confidence levels (i.e. the $\sim 1\mbox{-}\sigma$ level). Table~\ref{table:vesckSN} presents $v_{\rm esc}$ and $k_t$ (or $k$) derived for all the samples considered and curves shown in Fig.~\ref{fig:SNcombined}.

{\renewcommand{\arraystretch}{1.5}
\begin{table}
\caption{Escape velocity $(v_{\rm esc})$, power-law exponent $(k_t$, for the 5D and $k$ for the 6D samples) and the number of stars $(N_{\rm stars})$ for different distance estimates in the solar neighbourhood. The uncertainties in $v_{\rm esc}$ and $k_t$ (or $k$) are given by the marginalised $1\mbox{-}\sigma$ confidence levels.}     
\label{table:vesckSN}    
\centering                   
\begin{tabular}{c c c c}       
\hline\hline                 
Sample & $v_{\rm esc}$ (in km/s) & $k_t / k$ & $N_{\rm stars}$ \\     
\hline                        
   5D  & $497^{+8}_{-8}$ & $3.4^{+0.4}_{-0.3}$ & 10~000 \\
   6D+ & $497^{+12}_{-8}$ & $3.0^{+0.3}_{-0.2}$  & 2067\\
   6D ({\it Gaia} only) & $505^{+32}_{-16}$ & $3.0^{+0.7}_{-0.4}$ &  495 \\
\hline                                   
\end{tabular}
\end{table}
}

\section{Results: Beyond the solar neighbourhood}\label{sec:DR}

\begin{figure*}
    \centering
    \includegraphics[width=0.9\hsize]{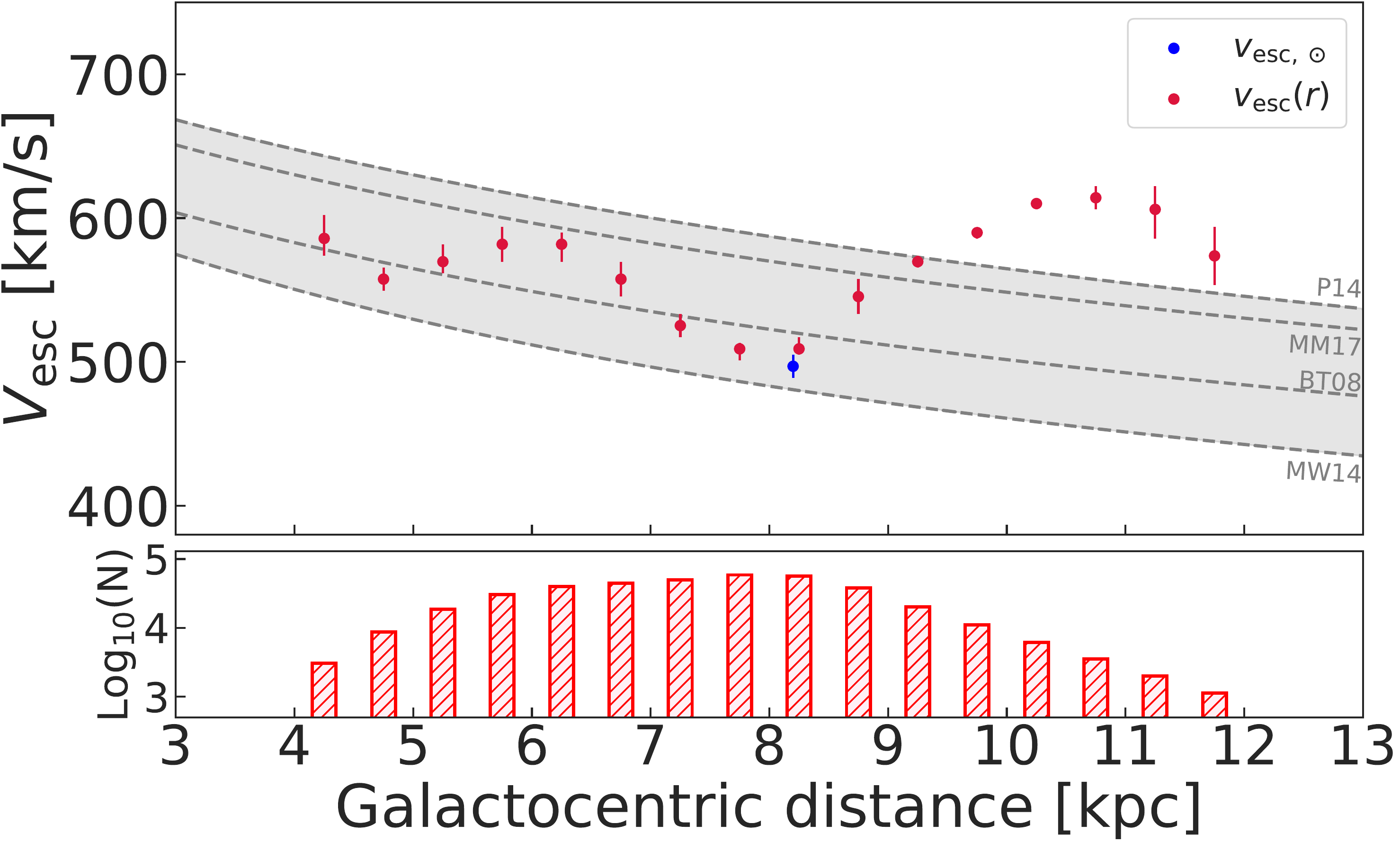}
    \caption{
    Escape velocity as a function of galactocentric distance (top). We also shown in grey the expected behaviour of the escape velocity for four often-used Milky Way models. The bottom panel shows the logarithm of the number of stars for each distance bin. The blue marker indicates the $v_{\rm esc}$ that we determined using a local sample of stars, see Sec.~\ref{sec:SN}. The error bars indicate the $1\mbox{-}\sigma$ confidence levels. We note that the local sample is not the same as the data sample at the bin of $\sim 8.2$~kpc, which is why the two markers there do not overlap exactly.}
    \label{fig:EvelDistanceRange}
\end{figure*}

\subsection{Determination of $v_{\rm esc}$}

We now proceed to determine $v_{\rm esc}$ as a function of
galactocentric distance. As we have seen in the Aurigaia halos, the
behaviour of $v_{\rm esc}$ as a function of distance can help in
identifying local `biases' or issues. We will here assume as prior
for $k_t$ the marginalised distribution obtained for the solar
neighbourhood $P(k_t)_{\rm SN}$, and shown in the right panel of
Fig.~\ref{fig:SNcombined} with the blue curve. Therefore, we implicitly
assume that $k_t$ remains constant over the distance range
probed. This assumption is justified by the Aurigaia simulations, as
shown in Fig.~\ref{fig:Aurigaia-combined}.

We sliced the data in $16$ concentric shells of $1$~kpc width, with $4<r<12$~kpc and centred on the Galactic Centre (as in Fig.~\ref{fig:DistRangeHists}). The number of stars per bin, with velocities larger than $v_{\rm cut}$, varies from $58~744$ to $1128$. In each shell, $v_{\rm cut}$ is determined adaptively by selecting the top 10~000 fastest stars. We do note that the results do not change significantly when the cut-off is fixed to $v_{\rm cut} = 250$ km/s.

Figure~\ref{fig:EvelDistanceRange} shows the trend of our estimate of $v_{\rm esc}$ with galactocentric distance. In each bin, the probability map is marginalised over the range $2.6<k_t<4.8$, after applying $P(k_t)_{\rm SN}$. This range in $k_t$ corresponds to the $3\mbox{-}\sigma$ interval of the posterior distribution of $P(k_t)_{\rm SN}$. We note that this is a very similar range to that assumed in S07. The use of the $P(k_t)_{\rm SN}$ prior beyond the solar neighbourhood has helped in the determination of $v_{\rm esc}$ for all the distance bins considered, despite the sometimes relatively small number of stars used. With the size of the samples presently available we could not have constrained both $k_t$ and $v_{\rm esc}$ simultaneously for all radial bins.

The behaviour of the escape velocity in the inner halo ($r < 8$ kpc) matches well 
the expectation from several Milky Way models. This can be seen by comparison to the predicted escape velocity plotted in the background of Fig.~\ref{fig:EvelDistanceRange} for the Piffl14, McMillan17, BT08 (model I), and MW14 potentials \citep[][all computed using the implementation from {\tt AGAMA}, \citealt{Vasiliev2019}]{Piffl2014, Mcmillan2017, Binney2008GalacticDynamics, Bovy2015Galpy:Dynamics}. The behaviour for the estimated $v_{\rm esc}$ shows small variations: a slight elevation at $\sim 6$~kpc and a dip at $\sim 4.5$~kpc, although it is fully consistent with a smooth increase towards the inner Galaxy. Furthermore, the amplitude of these variations is of a similar level as what we observed in the Aurigaia simulations. Curiously, our estimate of $v_{\rm esc}$ is higher outside of the solar radius (i.e. distance $> 8$ kpc). This cannot be driven by the mass profile of the Milky Way and can only mean that something is biasing the determination of $v_{\rm esc}$, as we discuss in detail next.

\subsection{The high $v_{\rm esc}$ outside of the solar radius}\label{sec:increase}

Several effects could lead to a higher $v_{\rm esc}$ outside the
solar radius, namely $(i)$ biases in the data (e.g. in the distance
estimate); $(ii)$ biases in the method (e.g. sample size), and $(iii)$
variations in the dynamical properties of the stars with distance. We
already explored the biases introduced by the first two categories in
Sec.~\ref{sec:data} (see also KH21) and
Sec.~\ref{sec:mockplaw}. Nevertheless, we also tested that when the
sample is downsized to a random subset of $5~000$ stars and bins
with fewer stars are excluded, the results do not change. We therefore now
focus on the third possibility: could the velocity distribution be different outside of the solar radius?

Careful inspection of Fig.~\ref{fig:DistRangeHists} shows that the velocity distribution is not contaminated by single outliers, even though in a relative sense (to the absolute number of objects) there seem to be more extreme velocity values in the outer radial bins. However, as mentioned earlier, the figure does show that the distributions seem to become more exponential with distance.

Curiously, we have seen a similar behaviour for Au-27 of the $v_{\rm esc}$-profile as observed for the 5D sample, see Fig.~\ref{fig:Aurigaia-combined}. In that case, the increase in $v_{\rm esc}$ was tentatively attributed to the presence of tidal debris (or at least lumpiness) moving with speeds close to the true escape velocity. 

\begin{figure}
    \centering
    \includegraphics[width=\hsize]{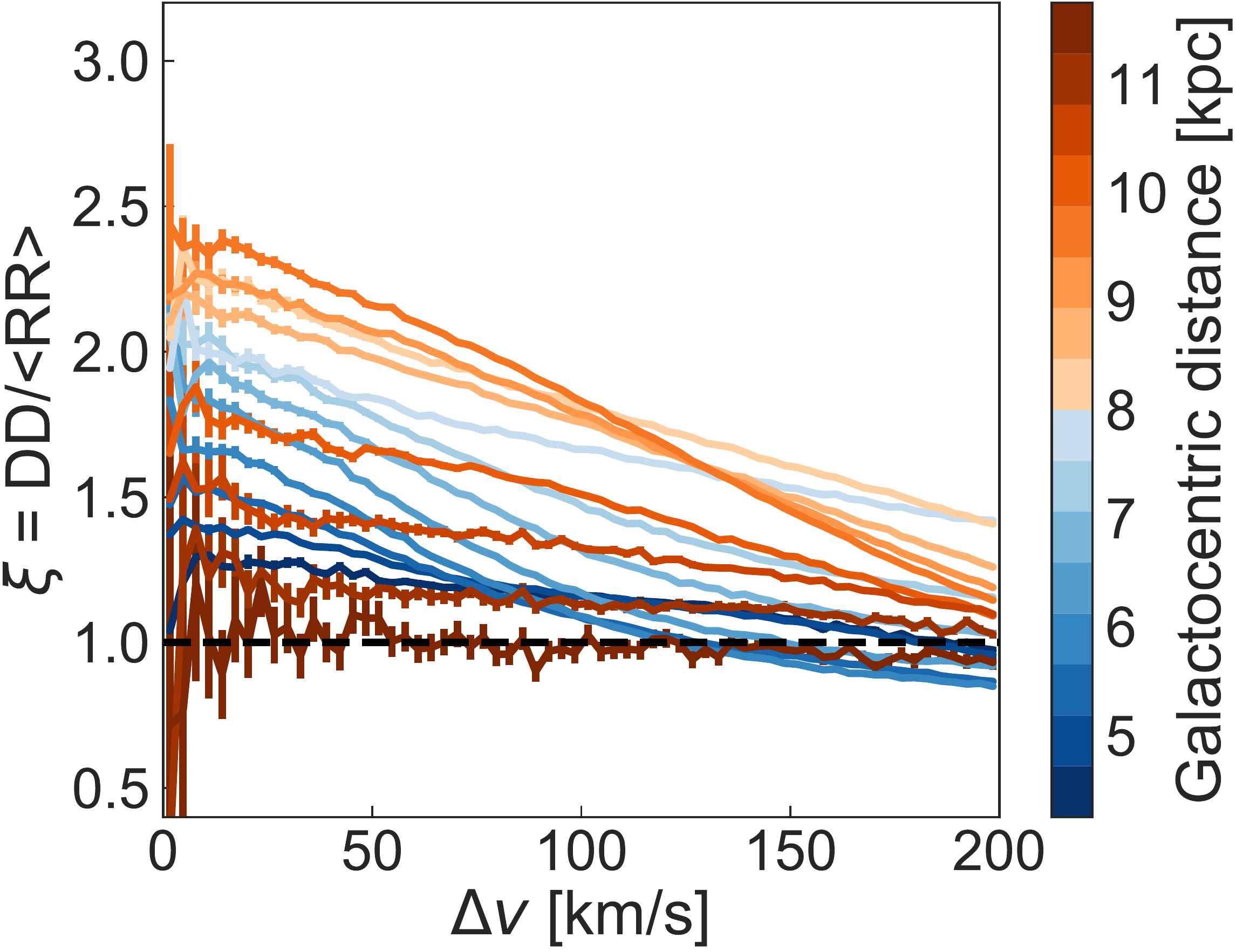}
    \caption{
    Two-point correlation function $\xi$ of the pseudo Cartesian velocities of the stars, binned by galactocentric distance. A correlation of $\xi>1$ indicates an excess of pairs compared to a random sample. The random sample $\langle RR \rangle$ is obtained by randomly shuffling the velocities in the galactic rest frame.}
    \label{fig:correlationfunc}
\end{figure}
 
With a two-point velocity correlation function, we test the statistical clustering of the stars in the tail of the velocity distribution of our 5D sample. An excess of pairs implies that the velocity distribution is not smooth. The two-point velocity correlation function is given by
\begin{equation}
    \xi(\Delta v) = \frac{DD (\Delta v)}{\langle RR (\Delta v)\rangle},
\end{equation}
where $DD(\Delta v)$ is the number of data-data pairs with a velocity separation of $\Delta v$ and similarly $\langle RR(\Delta v) \rangle$ is the mean number of random-random pairs obtained by randomly shuffling the velocities $100$ times. To this end, $v_\ell$ and $v_b$ are shuffled and the pseudo Cartesian velocities are re-calculated from Eqs.~\eqref{eq:vxyz} assuming $v_{\rm los} = 0$. Both the data and re-shuffled samples are cut-off at the velocity of the $10~000$th star, or 250 km/s if there are not enough stars per bin. Because of this re-sampling, most of the bins have an equal number of stars, except for those at large radii. 

Figure~\ref{fig:correlationfunc} shows the results of the correlation function $\xi$ for the 5D sample for the same distance bins as used throughout this paper.
The curves in Fig.~\ref{fig:correlationfunc} are coloured by the mean distance of the bin and the error bars show the uncertainty in $\xi$ estimated by the Poisson error in the number of counts per bin. A value $\xi=1$ indicates no excess correlation. We see that inside 
$8$ kpc, $\xi$ decreases with distance. Meanwhile, the bins just outside of this radius (light red) show the largest level of correlation over the full velocity range probed. The inner and outermost bins (dark colours) show the least correlation, although the uncertainties are large because of the low number of stars in these bins. There might also be an effect associated to the area of the shells increasing with distance squared, which results in the stars in the outer shells being physically more separated than those in the inner shells, and which could give rise to gradients in the trajectories of the stars and hence to lower correlation amplitudes.

The analysis of the velocity correlation function confirms that the properties of the velocity distribution change with distance. A hint of velocity clustering at $r \sim 10$~kpc in our 5D sample, similar (although of lower amplitude) to that seen for Au-27, could thus be responsible for this change.

\section{Discussion}\label{sec:discussion}

\subsection{Relating $v_{\rm esc}$ to the Milky Way's potential}\label{sec:methmassprof}

In Eq.~\eqref{eq:vescinfty} we defined the escape velocity as the velocity to reach  $r=\infty$. A more realistic definition is obtained by taking a different zero-point. No matter how one defines `escaping from the Milky Way', stars do not have to travel to infinity to be considered as escapees. For example, stars escaping to M31 make a much shorter journey $(r\approx800~{\rm kpc})$. 

Therefore, we use here the definition of P14, who take the escape velocity to be the velocity required to reach $3r_{340}$
\begin{equation}
    v_{\rm esc}(r\to 3r_{340}) \equiv \sqrt{2|\Phi(r)-\Phi(3r_{340})|},
    \label{eq:vesc}
\end{equation}
where $r_{340}$ is the radius within which the average halo density is 340$\times \rho_{\rm crit}$ (which is equal to $3H^2/8\pi G$, and where we assume $H = 73~{\rm km/s/Mpc}$). We note that this zero-point is set somewhat arbitrarily, the `true' value is directional dependent and might be a few km/s higher or lower. D19 use a different definition, which is for the star to escape to $2r_{200}~(\approx2.5r_{340})$. At the solar position, these two definitions result in a difference of $5~{\rm km/s}$. 

Because the potential is axisymmetric, $v_{\rm esc}$ varies as a function of cylindrical $R$ and $z$ for a fixed spherical $r$. In the plane of the disc, where the potential is the steepest, the escape velocity is the highest. Using the 
\cite{Mcmillan2017} potential, we estimate that $v_{\rm esc}$ decreases by $\sim 20~{\rm km/s}$ when moving $5~{\rm kpc}$ away from the plane of the disc, whereas at $10~{\rm kpc}$ the difference is about $50~{\rm km/s}$. 

To develop some intuition on how properties such as the mass of the Milky Way are related to $v_{\rm esc}$ we use the following equations. 
For a spherical potential, the gradient ${\rm d}v_{\rm esc}/{\rm d}r$ is related to the mass, circular velocity, and potential as
\begin{equation}
    \frac{{\rm d}\Phi(r)}{{\rm d}r} = 
    - v_{\rm esc}(r)\frac{{\rm d}v_{\rm esc}(r)}{{\rm d}r} = 
    \frac{v_{\rm circ}^2(r)}{r} = 
    \frac{GM(r)}{r^2}.
\end{equation}
Another insightful equation, given by Eq.~(2-22) of \cite{Binney1987GalacticDynamics} is
\begin{equation}
    v_{\rm esc}(r_\odot)^2 = 2v_{\rm circ}(r_\odot)^2 + 8\pi G\int_{r_\odot}^{\infty} r\rho (r)~\mathrm{d}r
    \label{eq:BT87}
\end{equation}
 (see S07). The circular velocity $v_{\rm circ}$ at the solar position is a direct measure of the mass {\it inside} of the solar radius. On the other hand, the escape velocity $v_{\rm esc}$ is a measure of the {\it total} gravitational potential. The two are related through a factor of $\sqrt{2}$ only if there is no mass outside of the radius where both are measured. In other words, the difference $v_{\rm esc}^2 - 2v_{\rm circ}^2$ at the solar neighbourhood probes the potential, and with it the mass distribution beyond the solar radius. 

\subsection{Estimating the mass of the Milky Way's halo}\label{sec:massestimate}

We will now use our
estimate of $v_{\rm esc}$
at the position of the Sun to
constrain
the mass of the halo of the Milky Way. The escape velocity and the gravitational potential of the Milky Way are related through Eq.~\eqref{eq:vesc}. A straightforward procedure to derive the mass of the Milky Way is to take an existing model and adjust the parameters of the halo such that it matches the $v_{\rm esc}$ measured for the solar neighbourhood. We follow closely the procedure outlined in Sec.~5 of D19, however here we will use the \cite{Mcmillan2017} potential and vary only the parameters of its dark halo, which is represented by an NFW profile \citep{Navarro1997AClustering}. 

The only issue with this procedure is that $v_{\rm esc}(r_\odot)$ is mostly sensitive to the mass \emph{outside} of the solar radius. As a result, 
fitting $v_{\rm esc}$ constrains only weakly the concentration of mass inside the solar radius. A solution is to use the circular velocity ($v_{\rm circ}$), which is sensitive to the mass \emph{inside} the solar radius, as an additional constraint. That is when fitting $v_{\rm esc}(r_\odot)$ we force the model to have a certain~ $v_{\rm circ}(r_\odot)$. 

\begin{figure}
    \centering
    \includegraphics[width=0.9\hsize]{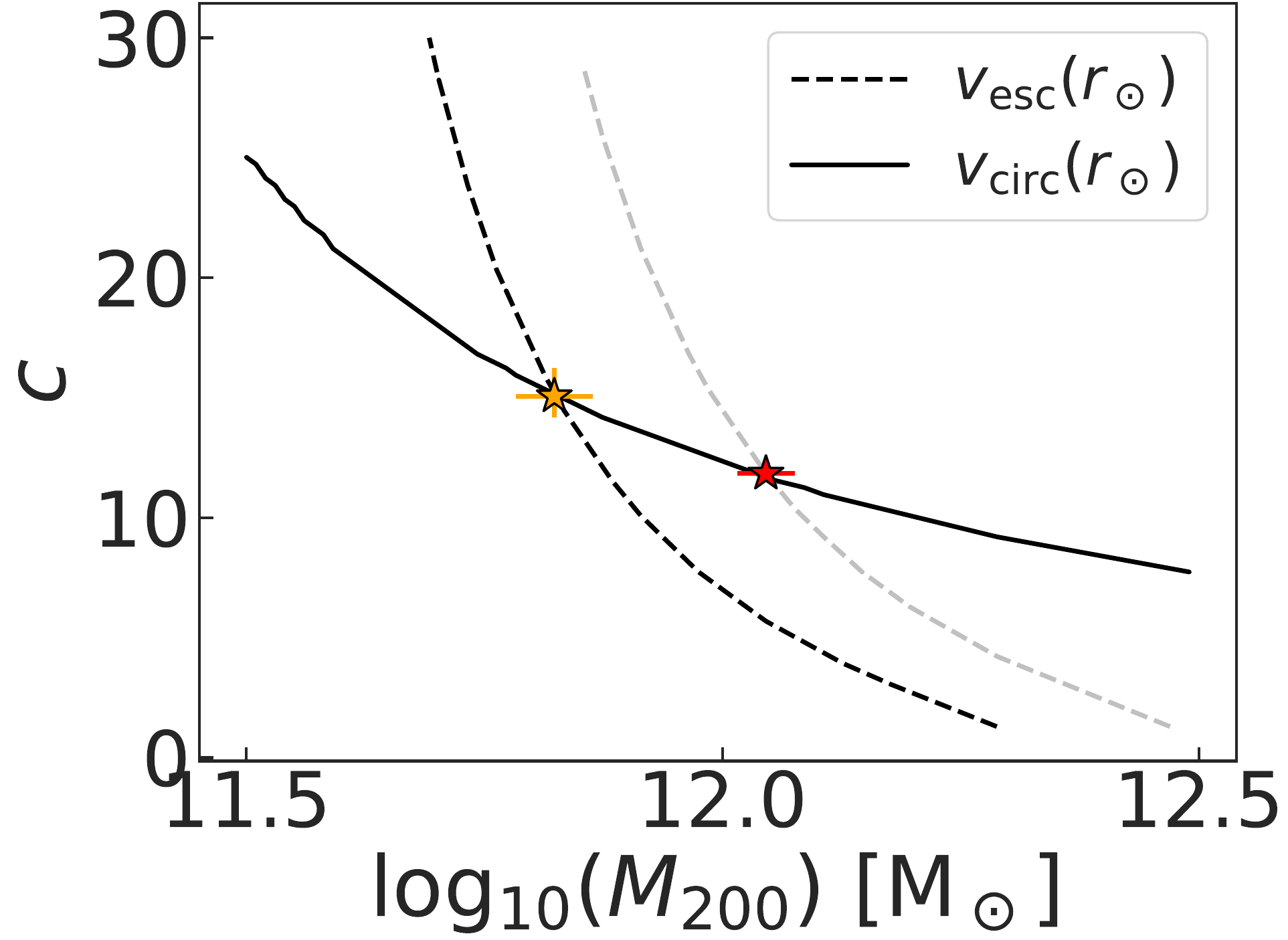}
    \caption{Best-fit combinations of the halo's mass and concentration parameter. The orange marker indicates the best fitting $M_{200}$ and $c$ parameters and the red marker shows the best fitting model after correcting $v_{\rm esc}$ for a 10\% offset. The error bars indicate the uncertainty in the halo parameters that is related to $1\mbox{-}\sigma$ variations in the estimate of $v_{\rm esc}$.}
    \label{fig:SNmass}
\end{figure}

The best-fitting potential is defined as the one that minimises 
\begin{equation}
    \eta = (v_{\rm esc}(r_\odot)-v_{\rm esc}^{\rm est})^2 +  (v_{\rm circ}(r_\odot) - 232.8~{\rm km/s})^2,
    \label{eq:massoptimiser}
\end{equation}
where we take $v_{\rm esc}^{\rm est}$ to be the maximum probability value found in the solar neighbourhood for the 5D sample (see Table~\ref{table:vesckSN}). The value for the circular velocity that we assume, $v_{\rm circ}(r_\odot) = 232.8~{\rm km/s}$, is the value 
that was used in the original \cite{Mcmillan2017} potential. We note that there is no freedom in choosing $v_{\rm circ}(r_\odot)$ because the data is only consistent with the value above, as it is used in the correction for the solar motion.

Figure~\ref{fig:SNmass} shows the values for Eq.~\eqref{eq:massoptimiser} for the ranges of $M_{200}$ and $c$ that we explore, namely $\log_{10}(M_{200})~[{\rm M}_\odot] \in [11.5,12.5]$ and $c \in [1,30]$. The solid line marks all models that have a correct $v_{\rm circ}(r_\odot)$ and the dashed lines mark all models that have the correct $v_{\rm esc}^{\rm est}$. The best fitting potential lies at the intersection of the two lines. The curves illustrate the benefit of including the $v_{\rm circ}$ in the fit. As we expected, the dashed curve is only weakly sensitive to $c$. The orange marker highlights the combination of $M_{200}$ and $c$ that best fits $v_{\rm esc}$ and $v_{\rm circ}$. Therefore the best fitting estimate of the mass is $M_{200} = 0.67^{+0.06}_{-0.06}\cdot10^{12}~{\rm M}_\odot$ and the corresponding concentration parameter is $c = 15.0^{+1.2}_{-0.9}$. The uncertainties are derived by calculating the best fitting $M_{200}$ and $c$ for the extreme cases of $v_{\rm esc}^{\rm est}+8$~km/s and $v_{\rm esc}^{\rm est}-8$~km/s, which are the limits given by the $1\mbox{-}\sigma$ level (e.g. Table~\ref{table:vesckSN}).

As we already mentioned, the LT90 method is likely to underestimate the $v_{\rm esc}$. Therefore, the mass and concentration parameters quoted above should be seen as a lower limit. As such it is consistent with the original potential of \cite{Mcmillan2017} in the sense that it is smaller and more concentrated. Moreover, this lower limit is also lower than most recent mass estimates \citep[c.f. Fig.~7][for a recent compilation]{Callingham2019TheDynamics}. 
If we now use the results from the analysis of the Aurigaia simulations and adjust for the $10\%$ underestimation of $v_{\rm esc}$ (grey dashed curve in Fig.~\ref{fig:SNmass}), we find that best-fit mass and concentration parameter are $M_{200} = 1.11^{+0.08}_{-0.07} \cdot10^{12} ~{\rm M}_\odot$ and $c = 11.8^{+0.3}_{-0.3}$.

\subsection{Stars that might be unbound}\label{sec:unbound}
A possibly interesting follow-up project is to measure the radial
velocities of the stars in the 5D sample that lie near the truncation
of the best-fit power-law. Using the maximum probability fit of the
velocity distribution we can calculate which stars have a high
probability of being unbound. Given the apparent $v^{\prime}_{t}$ and
its uncertainty, we can calculate the probability of these stars
having a true $v_t$ larger than $v_{\rm esc}$. We note that strictly
speaking the uncertainties are non-Gaussian, see also
Sec.~\ref{sec:method}. However, we assume that they are small
enough such that they may be approximated to~be~Gaussian.

For the set of stars that have apparent tangential velocities larger than the estimated $v_{\rm esc}$ we calculate the probability of the star being bound as
\begin{equation}
    P_{\rm bound} = \sum_{v_{e,i}} P_{\rm SN}(v_{e,i})\int_0^{v_{e,i}} f_G(v^{\prime}_t,v_t, \sigma_t) dv_t,
\end{equation}
where $P_{\rm SN}(v_e)$ is the posterior of $v_{\rm esc}$ marginalised over $k_t$ (i.e. the blue curve in the right panel of Fig.~\ref{fig:SNcombined}). The uncertainty distribution $f_G(v_t^\prime,v_t,\sigma_t)$ is defined such that it gives the probability of finding the star with a true velocity $v_t$ and uncertainty $\sigma_t$ with an apparent velocity in the range $(v^{\prime}_t,v^{\prime}_t + {\rm d}v^{\prime}_t)$. The probability of the star being unbound is simply $P_{\rm unbound} = 1 - P_{\rm bound}$. 

The list of sources that fall outside of the maximum probability value of $v_{\rm esc}$ is 
given in Table~\ref{table:unbound}. We stress that, very likely, the actual $v_{\rm esc}$ is higher than our best estimate. The values for $P_{\rm unbound}$ given here should therefore be considered as upper-limits. To emphasise this we also calculate the probability of these stars being unbound after correcting $v_{\rm esc}$ for a $10\%$ offset, based on our analysis in Sec.~\ref{sec:Aurigaia}. Only two sources remain unbound in this case, one of which just barely. The source with the largest probability of being unbound, with Gaia DR2 {\tt source\_id 2655054950237153664}, has been flagged in the {\tt faststars}\footnote{\url{https://faststars.space/}} database \citep{Guillochon2017AnData} as a potential hyper-velocity star. The source was first identified by \cite{Du2019NewDR2} based on its large tangential velocity.

About half the sources in Table~\ref{table:unbound} have an inward-pointing velocity vector, based on the pseudo velocities in the galactocentric frame. This makes it likely that the majority of these stars are bound to the Milky Way. Of course, there remains a possibility that the stars' velocity vectors point radially outwards when line-of-sight velocities are measured. However, for some stars the vectors will always point inwards even in the extreme case of $v_{\rm los} = \pm 500~{\rm km/s}$. One of such stars is the one with the highest probability of being unbound ({\tt source\_id 2655054950237153664}), which has a outward pointing velocity vector even for adopted line-of-sight velocities of $\pm 500$ km/s - and therefore might truly be unbound.

\begin{table}
\caption{High-velocity sources that are close to the escape velocity.}    

\label{table:unbound}    
\centering                   
\begin{tabular}{c c c c}       

\hline\hline \\[-0.35cm]             
{\tt source\_id} & $v^{\prime}_{\rm t}$ ($\epsilon_{v}$) & $P_{\rm unbound}$ & $P_{\rm unbound}^{+10\%}$ \\[0.1cm]    
\hline                        
5456509319663300096 & 499 (24) & 0.50  & 0.02 \\
4966291540726119936 & 500 (48) & 0.50  & 0.13 \\
1301068812277635968 & 500 (47) & 0.51  & 0.13 \\
3176805236597893248 & 502 (39) & 0.53  & 0.10 \\
600589157020469120  & 503 (50) & 0.53  & 0.15 \\
1300879558838744832 & 506 (41) & 0.57  & 0.13 \\
6669606511542374656 & 507 (61) & 0.55  & 0.22 \\
1268796702891972864 & 513 (46) & 0.61  & 0.19 \\
3289306720892701056 & 513 (64) & 0.58  & 0.26 \\
6843814817473042176 & 514 (35) & 0.65  & 0.14 \\
6587991790636824960 & 519 (36) & 0.70  & 0.17 \\
1142600839930233216 & 521 (65) & 0.63  & 0.30 \\
1948677828145591296 & 523 (57) & 0.66  & 0.29 \\
1831456179092459264 & 537 (43) & 0.80  & 0.35 \\
6085387089802067968 & 538 (63) & 0.73  & 0.40 \\
5845412900328041856 & 543 (57) & 0.77  & 0.42 \\
1981230244289202176 & 546 (48) & 0.83  & 0.43 \\
3495222399548253440 & 553 (68) & 0.78  & 0.49 \\
5175122643183339392 & 554 (30) & 0.96  & 0.49 \\
6270738976140076928 & 560 (49) & 0.88  & 0.54 \\
2655054950237153664 & 614 (62) & 0.96  & 0.82 \\
\hline                                   
\end{tabular}
\end{table}

\subsection{$v_{\rm esc}$ as tracer of the mass distribution}

\begin{figure}
    \centering
    \includegraphics[width=\hsize]{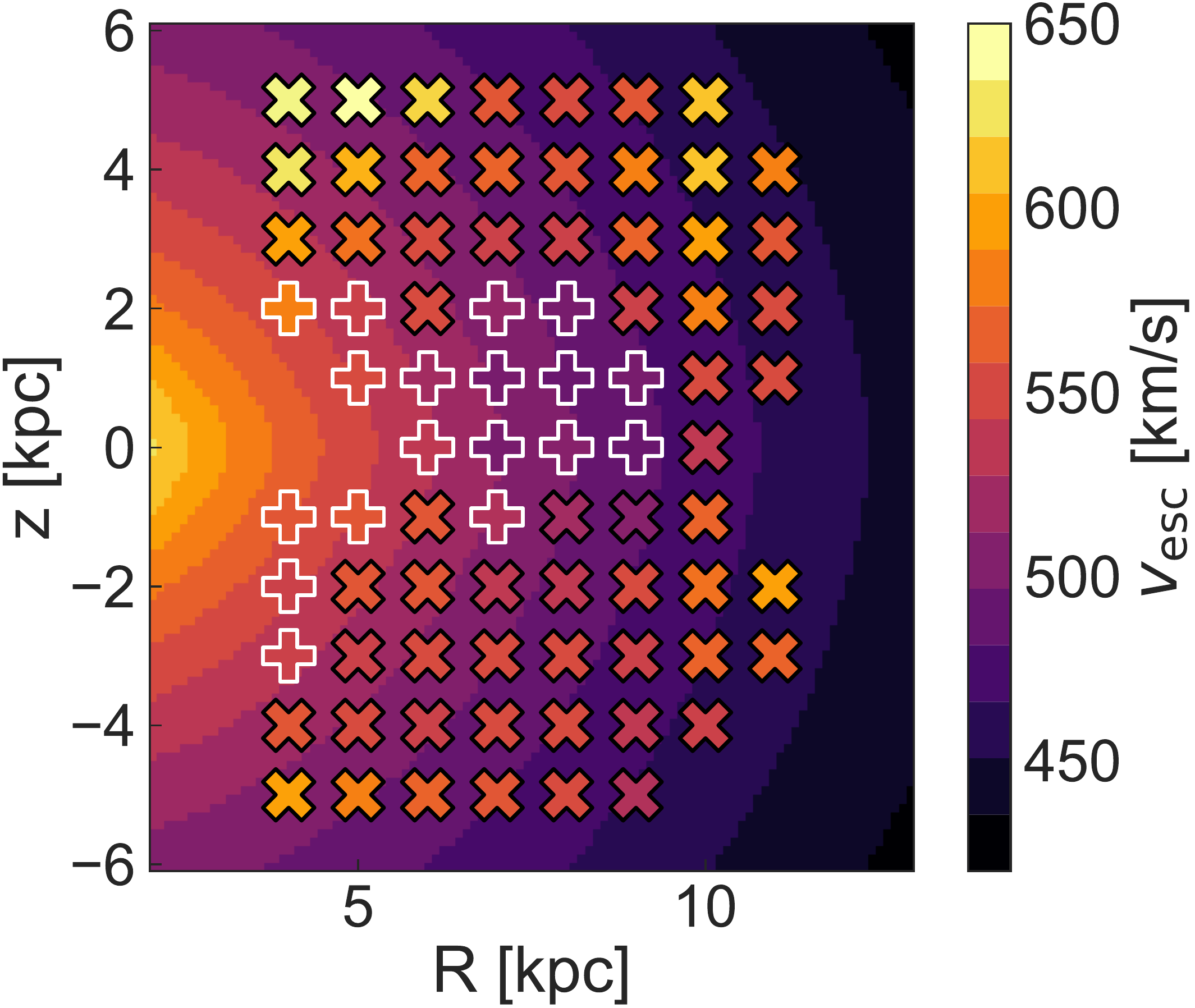}
    \caption{
    Escape velocity determined using the LT90 method in rings of constant cylindrical $R$ and $z$. The colour of the markers indicates the maximum-probability value for $v_{\rm esc}$. The colour map in the background shows expected isocontours for the escape velocity according to the McMillan17 potential, which assumes a spherical halo but whose parameters we have updated with the values from Sec.~\ref{sec:massestimate}. Volumes in which the $3\mbox{-}\sigma$ levels enclose the background-value are indicated with `+' markers, the `$\times$' markers indicate volumes where the determined $v_{\rm esc}$ is larger than expected.}
    \label{fig:vescRz}
\end{figure}

The luminous components of the Milky Way are most definitely not spherically symmetric. Because the escape velocity traces the potential we should ultimately measure it in axisymmetric coordinates rather than as a function of spherical radius. 
By estimating $v_{\rm esc}$ as a function of $z$ we can perhaps constrain the flattening of the halo, although with the current sample we are more sensitive to the contribution of the disc to the total potential of the Milky Way. Therefore, such an analysis would benefit from a large sample of stars probing deeper into the Milky Way's halo, such as what may become available with {\it Gaia}~(e)DR3.

Because of the large number of sources in our 5D sample, it is for the first time possible to explore the escape velocity as a function of cylindrical $R$ and $z$. We slice our 5D sample in overlapping bins of $8\times11$ volumes of $|R_c|<1$~kpc and $|z_c|<1$~kpc, where $R_c$ and $z_c$ are the centres of the volumes. Assuming that the Milky Way is perfectly axisymmetric, we include sources independent of their azimuthal angles. Bins with less than 500 stars are discarded. We use the same method to determine the escape velocity as we used in Sec.~\ref{sec:DR} and presented in Fig.~\ref{fig:EvelDistanceRange}. That is, we again assume the posterior distribution of $P_{\rm SN}(k_t)$ from the solar neighbourhood as prior on $k_t$. For computational reasons, we have decreased the size of the grid on which the probability distribution is evaluated to $50\times22$ points ranging from $400$ km/s $< v_{\rm esc}<$ $800$ km/s and $2.6<k_t<4.8$ (which corresponds to the $3\mbox{-}\sigma$ levels in the solar neighbourhood).

Figure~\ref{fig:vescRz} shows the escape velocity in these volumes
(coloured, large markers) with a colour map corresponding to the
escape velocity predicted by the McMillan17 model, with the updated
lower limit of the halo mass computed in
Sec.~\ref{sec:massestimate}. Therefore this model is based on the
estimate for $v_{\rm esc}$ that is biased low and we use it to predict what this
estimate would be at other locations for a spherical NFW halo. The large
`+' markers in Fig.~\ref{fig:vescRz} indicate volumes in which the
$3\mbox{-}\sigma$ levels of the $v_{\rm esc}$ include the expected value. The large
`$\times$' markers indicate volumes where the $v_{\rm esc}$ expected
from the updated McMillan17 potential lies outside of the $3\mbox{-}\sigma$
confidence level of the maximum probability determination of
$v_{\rm esc}$. Interestingly, the distribution is not fully symmetric
in $z$. The fact that $v_{\rm esc}$ does not match the expected value
in many locations could potentially indicate a bias in the estimated
$v_{\rm esc}$ at the solar neighbourhood. Another possibility is that
the decrease in the strength of the potential with $z$ is less steep
than expected for a spherical halo (e.g. pointing to a prolate halo or
less strong influence from the disc).

\section{Conclusions}\label{sec:conclusions}

We used a sample of halo stars with large tangential velocities
to constrain the escape velocity in the vicinity of the Sun and as a
function of galactocentric distance. We applied the well-known
LT90 method, which fits the high-velocity tail (i.e. above some
velocity $v_{\rm cut}$) of the velocity distribution with a power law
of the form $(v_{\rm esc} - v)^k$. In the process of applying the
method, we identified a number of shortcomings.

The study presented here constitutes the first application of the
method to a sample of stars using tangential velocities only. We have
found that in practice, the estimated value for the parameter $k$ is
not exactly what is predicted by LT90 (namely $k_t = k$), except
really in the tail of the distribution, in the limit where $v_{\rm cut}$
differs by 10\% from $v_{\rm esc}$. Unfortunately, the value of
$v_{\rm cut}$ typically chosen is farther away from $v_{\rm esc}$
because enough stars ($\sim 10^4$) with high velocity
need to be present in the sample for a precise estimate
of $v_{\rm esc}$. A similar conclusion may be reached when applying
the method to radial velocity samples. Therefore, care is necessary
when comparing the values of
$k$ for different studies in the literature. Fortunately,
$v_{\rm esc}$ is unaffected.

In addition, and as previously discussed in the literature, the
$v_{\rm esc}$ determined via the LT90 method is most likely a
\emph{lower limit}. To get a handle on this bias we have tested the
method on two mock {\it Gaia} catalogues from the Aurigaia project
\citep{Grand2018Aurigaia:Simulations}. In these simulated galaxies the
estimated $v_{\rm esc}$ are $\sim10\%$ lower than the true values,
close to the 7\% bias found in a similar study by
\cite{Grand2019TheHalo}. Based on this result, when reporting our
estimates of the escape velocity, we also quote the value obtained by
applying a 10\% correction. However, we note that there is no
guarantee that the Milky Way's halo is truncated at a similar level as
the Aurigaia halos. The truncation of the velocity distribution will
be dependent on the (recent) assembly history of the Galaxy and for
the simulations, it might depend on the numerical resolution.

In the solar neighbourhood, using a 5D sample, we determine a very precise
estimate of the escape velocity, $v_{\rm esc} = 497^{+8}_{-8}~{\rm km/s}$, and
power-law exponent $k_t = 3.4^{+0.4}_{-0.3}$. The quoted uncertainties are given
by the level where the
probability has dropped to $61\%$ of the maximum value (i.e. the
$\sim 1\mbox{-}\sigma$ level). These values agree well with previous works,
but this is the first time, we can determine (a lower limit to) the
escape velocity with such high confidence. This value for
$v_{\rm esc}$ agrees remarkably well that obtained when we use a local
sample of halo stars with full phase-space information. Applying the
10\% fix would mean that the true escape velocity is 
$v_{\rm esc}^{+10\%} = 552~{\rm km/s}$.

We also determine $v_{\rm esc}$ as a function of galactocentric
distance. We find that the escape velocity is larger in the inner
halo than at the solar radius. This matches well the behaviour expected
from smooth Milky Way models. Unexpectedly, for radii beyond
$8$ kpc, $v_{\rm esc}$ is also higher than
at the solar radius (see Fig.~\ref{fig:DistRangeHists}).
Hints of a similar trend were picked up by M18, but at a
much lower significance level because of their limited sample size.

Interestingly, we find that the behaviour of $v_{\rm esc}$ outside of
the solar radius is paired with a change of shape of the velocity
distribution. For example, the tail of the velocity distribution
becomes more exponential (and less power law-like) with galactocentric
distance (see Fig.~\ref{fig:DistRangeHists}). Also, the velocities in
the bins outside of $8$~kpc show a higher degree of correlation as
measured by the velocity correlation function. Therefore, we conclude
that the bump in $v_{\rm esc}$ in the outskirts is likely
driven by a change in the kinematic properties of the sample as a
function of galactocentric distance. Coincidentally, we found a
similar effect in one of the Aurigaia halos analysed, where a velocity
bump (presumably related to a clump or a non-phase-mixed structure in
the halo) dominates the tail near the escape velocity.

The estimated $v_{\rm esc}$ can be used to provide a very precise
estimate of the mass of the halo of the Milky Way. To this end, we
have adjusted the halo component of the \cite{Mcmillan2017} Milky Way
potential (which is a spherical NFW profile), while keeping the
other components fixed. The halo
parameters that best fit the estimated $v_{\rm esc}(r_\odot)$ are
$M_{200} = 0.67^{+0.06}_{-0.06}\cdot10^{12}~{\rm M}_\odot$ and
$c = 15^{+1.2}_{-0.9}$, where we used $v_{\rm circ}(r_\odot)$ as an
additional constraint. When we apply the tentative $10\%$-fix we find
that the best fitting halo has $M_{200}^{+10\%} = 1.11^{+0.08}_{-0.07}
\cdot10^{12} ~{\rm M}_\odot$ and $c^{+10\%} = 11.8^{+0.3}_{-0.3}$.

The method to determine $v_{\rm esc}$ consists in fitting the tail of
the velocity distribution with a parametrised model. Using the best
fitting model obtained, we can also establish if there are any unbound
stars in the solar neighbourhood. That is, we may calculate which
stars have a high probability of having a true velocity that is larger
than the determined escape velocity. We list these stars in
Table~\ref{table:unbound}. Their pseudo velocities (without
the line-of-sight velocity), however, suggest they are not all
unbound: their velocity vectors point both inwards and outwards. If
these high-velocity stars were truly escaping we would expect them to
all be on radially outbound trajectories. Nonetheless, it might be
interesting to follow-up these stars. When taking into account the
tentative $10\%$-fix only one candidate with a large probability of
being unbound remains: {\it Gaia} DR2 {\tt source\_id 
2655054950237153664}. This star was first flagged as being unbound
by \mbox{\cite{Du2019NewDR2}}.

Finally, we discuss a tentative method to probe the mass distribution
of the Milky Way by determining $v_{\rm esc}$ as a function of
$(R,z)$. We find that escape velocity values that are weakly asymmetric with
respect to the galactic plane, and also tentative indication that
the halo may be prolate. However, for more robust conclusions a
larger sample with more accurate distances and that probes deeper into
the Milky Way is necessary. We hope that such a sample will become
available with {\it Gaia}~(e)DR3.

\begin{acknowledgements}
We gratefully acknowledge financial support from a VICI grant and a Spinoza Prize from the Netherlands Organisation for Scientific Research (NWO) and HHK is grateful for the support from the Martin A. and Helen Chooljian Membership at the Institute for Advanced Study. HHK thanks Daniel Foreman-Mackey, Scott Tremaine, and Rosemary Wyse for stimulating discussions on an early version of this work, that took place during the KITP Santa Barbara long-term program `Dynamical Models for Stars and Gas in Galaxies in the Gaia Era', which was supported in part by the National Science Foundation under Grant No. NSF PHY-1748958. This work has made use of data from the European Space Agency (ESA) mission Gaia (\url{http://www.cosmos.esa.int/gaia}), processed by the Gaia Data Processing and Analysis Consortium (DPAC, \url{http://www.cosmos.esa.int/web/gaia/dpac/consortium}). Funding for the DPAC has been provided by national institutions, in particular the institutions participating in the Gaia Multilateral Agreement. 
For the analysis, the following software packages have been used: {\tt vaex} \citep{Breddels2018}, {\tt scipy} \citep{Virtanen2020SciPyPython}, {\tt numpy} \citep{VanDerWalt2011TheComputation}, {\tt matplotlib} \citep{Hunter2007Matplotlib:Environment}, {\tt jupyter notebooks} \citep{Kluyver2016JupyterWorkflows}, and {\tt Mathematica} \citep{Mathematica}.
\end{acknowledgements}

\bibliographystyle{aa} 
\bibliography{EscVel}

\begin{thebibliography}{65}
\expandafter\ifx\csname natexlab\endcsname\relax\def\natexlab#1{#1}\fi

\bibitem[{Abolfathi {et~al.}(2018)Abolfathi, Aguado, Aguilar, Prieto, Almeida,
  Ananna, Anders, Anderson, Andrews, Anguiano, Arag{\'{o}}n-Salamanca,
  Argudo-Fern{\'{a}}ndez, Armengaud, Ata, Aubourg, Avila-Reese, Badenes,
  Bailey, Balland, Barger, Barrera-Ballesteros, Bartosz, Bastien, Bates,
  Baumgarten, Bautista, Beaton, Beers, Belfiore, Bender, Bernardi, Bershady,
  Beutler, Bird, Bizyaev, Blanc, Blanton, Blomqvist, Bolton, Boquien,
  Borissova, Bovy, Bradna~Diaz, Nielsen~Brandt, Brinkmann, Brownstein, Bundy,
  Burgasser, Burtin, Busca, Ca{\~{n}}as, Cano-D{\'{i}}az, Cappellari, Carrera,
  Casey, Sodi, Chen, Cherinka, Chiappini, Choi, Chojnowski, Chuang, Chung,
  Clerc, Cohen, Comerford, Comparat, do~Nascimento, da~Costa, Cousinou, Covey,
  Crane, Cruz-Gonzalez, Cunha, Ilha, Damke, Darling, Davidson, Dawson,
  de~Icaza~Lizaola, Macorra, de~la Torre, De~Lee, Sainte~Agathe,
  Deconto~Machado, Dell’Agli, Delubac, Diamond-Stanic, Donor, Downes, Drory,
  Mas~des Bourboux, Duckworth, Dwelly, Dyer, Ebelke, Eigenbrot, Eisenstein,
  Elsworth, Emsellem, Eracleous, Erfanianfar, Escoffier, Fan, Alvar,
  Fernandez-Trincado, Cirolini, Feuillet, Finoguenov, Fleming, Font-Ribera,
  Freischlad, Frinchaboy, Fu, Chew, Galbany, Garc{\'{i}}a~P{\'{e}}rez,
  Garcia-Dias, Garc{\'{i}}a-Hern{\'{a}}ndez, Garma~Oehmichen, Gaulme, Gelfand,
  Gil-Mar{\'{i}}n, Gillespie, Goddard, Gonz{\'{a}}lez~Hern{\'{a}}ndez,
  Gonzalez-Perez, Grabowski, Green, Grier, Gueguen, Guo, Guy, Hagen, Hall,
  Harding, Hasselquist, Hawley, Hayes, Hearty, Hekker, Hernandez,
  Hernandez~Toledo, Hogg, Holley-Bockelmann, Holtzman, Hou, Hsieh, Hunt,
  Hutchinson, Hwang, Jimenez~Angel, Johnson, Jones, J{\"{o}}nsson, Jullo,
  Sakil~Khan, Kinemuchi, Kirkby, Kirkpatrick~IV, Kitaura, Knapp, Kneib,
  Kollmeier, Lacerna, Lane, Lang, Law, Le~Goff, Lee, Li, Li, Lian, Liang, Lima,
  Lin, Long, Lucatello, Lundgren, Mackereth, MacLeod, Mahadevan, Geimba~Maia,
  Majewski, Manchado, Maraston, Mariappan, Marques-Chaves, Masseron, Masters,
  McDermid, McGreer, Melendez, Meneses-Goytia, Merloni, Merrifield, Meszaros,
  Meza, Minchev, Minniti, Mueller, Muller-Sanchez, Muna, Mu{\~{n}}oz, Myers,
  Nair, Nandra, Ness, Newman, Nichol, Nidever, Nitschelm, Noterdaeme,
  O’Connell, Oelkers, Oravetz, Oravetz, Ort{\'{i}}z, Osorio, Pace, Padilla,
  Palanque-Delabrouille, Palicio, Pan, Pan, Parikh, P{\^{a}}ris, Park, Peirani,
  Pellejero-Ibanez, Penny, Percival, Perez-Fournon, Petitjean, Pieri,
  Pinsonneault, Pisani, Prada, Prakash, de~Andrade~Queiroz, Raddick, Raichoor,
  Rembold, Richstein, Riffel, Riffel, Rix, Robin, Torres,
  Rom{\'{a}}n-Z{\'{u}}{\~{n}}iga, Ross, Rossi, Ruan, Ruggeri, Ruiz, Salvato,
  S{\'{a}}nchez, S{\'{a}}nchez, Almeida, S{\'{a}}nchez-Gallego, Rojas,
  Santiago, Schiavon, Schimoia, Schlafly, Schlegel, Schneider, Schuster,
  Schwope, Seo, Serenelli, Shen, Shen, Shetrone, Shull, Aguirre, Simon,
  Skrutskie, Slosar, Smethurst, Smith, Sobeck, Somers, Souter, Souto, Spindler,
  Stark, Stassun, Steinmetz, Stello, Storchi-Bergmann, Streblyanska,
  Stringfellow, Su{\'{a}}rez, Sun, Szigeti, Taghizadeh-Popp, Talbot, Tang, Tao,
  Tayar, Tembe, Teske, Thakar, Thomas, Tissera, Tojeiro, Tremonti, Troup, Urry,
  Valenzuela, Bosch, Vargas-Gonz{\'{a}}lez, Vargas-Maga{\~{n}}a, Vazquez,
  Villanova, Vogt, Wake, Wang, Weaver, Weijmans, Weinberg, Westfall, Whelan,
  Wilcots, Wild, Williams, Wilson, Wood-Vasey, Wylezalek, Xiao, Yan, Yang,
  Ybarra, Y{\`{e}}che, Zakamska, Zamora, Zarrouk, Zasowski, Zhang, Zhao, Zhao,
  Zheng, Zheng, Zhou, Zhu, Zinn, \& Zou}]{Abolfathi2018TheExperiment}
Abolfathi, B., Aguado, D.~S., Aguilar, G., {et~al.} 2018, \apjs, 235, 42

\bibitem[{Aguilar \& White(1986)}]{Aguilar1986THEGALAXIES}
Aguilar, L.~A. \& White, S. D.~M. 1986, \apj, 307, 97

\bibitem[{Beckmann(1962)}]{Beckmann1962}
Beckmann, P. 1962, Journal of Research of the National Bureau of Standards,
  Section D: Radio Propagation, 66D, 231

\bibitem[{Belokurov {et~al.}(2018)Belokurov, Erkal, Evans, Koposov, \&
  Deason}]{Belokurov2018Co-formationHalo}
Belokurov, V., Erkal, D., Evans, N.~W., Koposov, S.~E., \& Deason, A.~J. 2018,
  \mnras, 478, 611

\bibitem[{Binney \& Tremaine(1987)}]{Binney1987GalacticDynamics}
Binney, J. \& Tremaine, S. 1987, {Galactic dynamics}, v1 edn. (Princeton
  University Press)

\bibitem[{Binney \& Tremaine(2008)}]{Binney2008GalacticDynamics}
Binney, J. \& Tremaine, S. 2008, {Galactic dynamics} (Princeton University
  Press), 885

\bibitem[{Boubert {et~al.}(2018)Boubert, Guillochon, Hawkins, Ginsburg, Evans,
  \& Strader}]{Boubert2018}
Boubert, D., Guillochon, J., Hawkins, K., {et~al.} 2018, \mnras, 479, 2789

\bibitem[{Boubert {et~al.}(2019)Boubert, Strader, Aguado, Seabroke, Koposov,
  Sanders, Swihart, Chomiuk, \& Evans}]{Boubert2019}
Boubert, D., Strader, J., Aguado, D., {et~al.} 2019, \mnras, 486, 2618

\bibitem[{Bovy(2015)}]{Bovy2015Galpy:Dynamics}
Bovy, J. 2015, \apjs, 216, 29

\bibitem[{Breddels \& Veljanoski(2018)}]{Breddels2018}
Breddels, M.~A. \& Veljanoski, J. 2018, \aap, 618, 13

\bibitem[{Brown(2015)}]{Brown2015HypervelocityStars}
Brown, W.~R. 2015, \araa, 53, 15

\bibitem[{Callingham {et~al.}(2019)Callingham, Cautun, Deason, Frenk, Wang,
  G{\'{o}}mez, Grand, Marinacci, \& Pakmor}]{Callingham2019TheDynamics}
Callingham, T.~M., Cautun, M., Deason, A.~J., {et~al.} 2019, \mnras, 484, 5453

\bibitem[{Chan \& Bovy(2020)}]{Chan2019TheStars}
Chan, V.~C. \& Bovy, J. 2020, \mnras, 493, 4367

\bibitem[{Cui {et~al.}(2012)Cui, Zhao, Chu, Li, Li, Zhang, Su, Yao, Wang, Xing,
  Li, Zhu, Wang, Gu, Luo, Xu, Zhang, Liu, Zhang, Yang, Cao, Chen, Chen, Chen,
  Chen, Chu, Feng, Gong, Hou, Hu, Hu, Hu, Jia, Jiang, Jiang, Jiang, Jin, Li,
  Li, Li, Liu, Liu, Lu, Mao, Men, Qi, Qi, Shi, Tang, Tao, Wang, Wang, Wang,
  Wang, Wang, Wang, Wang, Wang, Wang, Wang, Wang, Wang, Xu, Xu, Yang, Yu, Yuan,
  Yuan, Zhai, Zhang, Zhang, Zhang, Zhao, Zhou, Zhou, Zhu, \& Zou}]{Cui2012}
Cui, X.-Q., Zhao, Y.-H., Chu, Y.-Q., {et~al.} 2012, Research in Astronomy and
  Astrophysics, 12, 1197

\bibitem[{Deason {et~al.}(2012)Deason, Belokurov, Evans, Koposov, Cooke,
  Pe{\~{n}}arrubia, Laporte, Fellhauer, Walker, \&
  Olszewski}]{Deason2012TheHalo}
Deason, A.~J., Belokurov, V., Evans, N.~W., {et~al.} 2012, \mnras, 425, 2840

\bibitem[{Deason {et~al.}(2019)Deason, Fattahi, Belokurov, Wyn~Evans, Grand,
  Marinacci, \& Pakmor}]{Deason2019TheSpeed}
Deason, A.~J., Fattahi, A., Belokurov, V., {et~al.} 2019, \mnras, 485, 3514

\bibitem[{Dierickx \& Loeb(2017)}]{Dierickx2017a}
Dierickx, M. I.~P. \& Loeb, A. 2017, \apj, 847, 42

\bibitem[{Du {et~al.}(2019)Du, Li, Yan, Newberg, Shi, Ma, Chen, \&
  Wu}]{Du2019NewDR2}
Du, C., Li, H., Yan, Y., {et~al.} 2019, \apjs, 244, 4

\bibitem[{Eilers {et~al.}(2019)Eilers, Hogg, Rix, \& Ness}]{Eilers2019TheKpc}
Eilers, A.-C., Hogg, D.~W., Rix, H.-W., \& Ness, M.~K. 2019, \apj, 871, 120

\bibitem[{Fragione \& Loeb(2017)}]{Fragione2016}
Fragione, G. \& Loeb, A. 2017, New Astronomy, 55, 32

\bibitem[{Fritz {et~al.}(2020)Fritz, {Di Cintio}, Battaglia, Brook, \&
  Taibi}]{Fritz2020}
Fritz, T.~K., {Di Cintio}, A., Battaglia, G., Brook, C., \& Taibi, S. 2020,
  \mnras, 494, 5178

\bibitem[{{Gaia Collaboration} {et~al.}(2016){Gaia Collaboration}, Prusti, J~de
  Bruijne, A~Brown, Vallenari, Babusiaux, L~Bailer-Jones, Bastian, Biermann,
  Evans, Eyer, Jansen, Jordi, Klioner, Lammers, Lindegren, Luri, Mignard,
  Milligan, Panem, Poinsignon, Pourbaix, Randich, Sarri, Sartoretti, Siddiqui,
  Soubiran, Valette, van Leeuwen, Walton, Aerts, Arenou, Cropper, Drimmel,
  H{\o}g, Katz, Lattanzi, Grebel, Holland, Huc, Passot, Bramante, Cacciari,
  Casta{\~{n}}eda, Chaoul, Cheek, De~Angeli, Fabricius, Guerra,
  Hern{\'{a}}ndez, Jean-Antoine-Piccolo, Masana, Messineo, Mowlavi,
  Nienartowicz, Ord{\'{o}}{\~{n}}ez-Blanco, Panuzzo, Portell, Richards,
  Barache, Barata, Barbier, Barblan, Baroni, Barrado~Navascu{\'{e}}s, Barros,
  Barstow, Becciani, Bellazzini, Bellei, Bello~Garc{\'{i}}a, Belokurov,
  Bendjoya, Berihuete, Bianchi, Bienaym{\'{e}}, Billebaud, Blagorodnova,
  Blanco-Cuaresma, Boch, Bombrun, Borrachero, Bouquillon, Bourda, Bouy,
  Bragaglia, Breddels, Brouillet, Br{\"{u}}semeister, Bucciarelli, Budnik,
  Burgess, Burgon, Burlacu, Busonero, Buzzi, Caffau, Cambras, Campbell,
  Cancelliere, Cantat-Gaudin, Carlucci, Carrasco, Castellani, Charlot, Charnas,
  Charvet, Chassat, Chiavassa, Clotet, Cocozza, Collins, Collins, Costigan,
  Crifo, G~Cross, Crosta, Crowley, Dafonte, Damerdji, Dapergolas, David, David,
  De~Cat, de~Felice, de~Laverny, De~Luise, De~March, de~Martino, de~Souza,
  Debosscher, del Pozo, Delbo, Delgado, Delgado, di~Marco, Di~Matteo, Diakite,
  Distefano, Dolding, Dos~Anjos, Drazinos, Dur{\'{a}}n, Dzigan, Ecale,
  Edvardsson, Enke, Erdmann, Escolar, Espina, Evans, Eynard~Bontemps, Fabre,
  Fabrizio, Faigler, Falc{\~{a}}o, Farr{\`{a}}s~Casas, Faye, Federici,
  Fedorets, Fern{\'{a}}ndez-Hern{\'{a}}ndez, Fernique, Fienga, Figueras,
  Filippi, Findeisen, Fonti, Fouesneau, Fraile, Fraser, Fuchs, Furnell, Gai,
  Galleti, Galluccio, Garabato, Garc{\'{i}}a-Sedano, Gar{\'{e}}, Garofalo,
  Garralda, Gavras, Gerssen, Geyer, Gilmore, Girona, Giuffrida, Gomes,
  Gonz{\'{a}}lez-Marcos, Gonz{\'{a}}lez-N{\'{u}}{\~{n}}ez,
  Gonz{\'{a}}lez-Vidal, Granvik, Guerrier, Guillout, Guiraud, G{\'{u}}rpide,
  Guti{\'{e}}rrez-S{\'{a}}nchez, Guy, Haigron, Hatzidimitriou, Haywood, Heiter,
  Helmi, Hobbs, Hofmann, Holl, Holland, S~Hunt, Hypki, Koubsky, Kowalczyk,
  Krone-Martins, Kudryashova, Kull, Bachchan, Lacoste-Seris, Lanza, Lavigne,
  Le~Poncin-Lafitte, Lebreton, Lebzelter, Leccia, Leclerc, Lecoeur-Taibi,
  Lemaitre, Lenhardt, Leroux, Liao, Licata, P~Lindstr{\o}m, Lister, Livanou,
  Lobel, L{\"{o}}ffler, L{\'{o}}pez, Lopez-Lozano, Lorenz, Loureiro, MacDonald,
  Magalh{\~{a}}es~Fernandes, Managau, Mann, Mantelet, Marchal, Pagani, Pagano,
  Pailler, Palacin, Palaversa, Parsons, Paulsen, Pecoraro, Pedrosa,
  Pentik{\"{a}}inen, Pereira, Pichon, Piersimoni, Pineau, Plachy, Plum,
  Poujoulet, Pr{\v{s}}a, Pulone, Ragaini, Rago, Rambaux, Ramos-Lerate, Ranalli,
  Rauw, Read, Regibo, Renk, Reyl{\'{e}}, Ribeiro, Rimoldini, Ripepi, Riva,
  Rixon, Roelens, Romero-G{\'{o}}mez, Rowell, Royer, Rudolph, Ruiz-Dern,
  Sadowski, Sagrist{\`{a}}~Sell{\'{e}}s, Sahlmann, Salgado, Salguero, Sarasso,
  Savietto, Schnorhk, Schultheis, Sciacca, Segol, Segovia, Segransan, Serpell,
  Shih, Smareglia, Smart, Smith, Solano, Solitro, Sordo, Soria~Nieto, Souchay,
  Spagna, Spoto, Stampa, Steele, Steidelm{\"{u}}ller, Stephenson, Stoev, Suess,
  S{\"{u}}veges, Surdej, Szabados, Szegedi-Elek, Tapiador, Taris, Tauran,
  Taylor, Teixeira, Terrett, Tingley, Trager, Turon, Ulla, Utrilla, Valentini,
  van Elteren, Van~Hemelryck, van Leeuwen, Varadi, Vecchiato, Veljanoski, Via,
  Vicente, Vogt, Voss, Votruba, Voutsinas, Walmsley, Weiler, Weingrill, Werner,
  Wevers, Whitehead, Wyrzykowski, Yoldas, {\v{Z}}erjal, Zucker, Zurbach,
  Zwitter, Alecu, Allen, Allende~Prieto, Amorim, Anglada-Escud{\'{e}},
  Arsenijevic, Azaz, Balm, Beck, Bernstein, Bigot, Bijaoui, Blasco, Bonfigli,
  Bono, Boudreault, Bressan, Brown, Brunet, Bunclark, Buonanno, Butkevich,
  Carret, Carrion, Chemin, Ch{\'{e}}reau, Corcione, Darmigny, de~Boer,
  de~Teodoro, de~Zeeuw, Delle~Luche, Domingues, Dubath, Fodor, Fr{\'{e}}zouls,
  Fries, Fustes, Fyfe, Gallardo, Gallegos, Gardiol, Gebran, Gomboc,
  G{\'{o}}mez, Grux, Gueguen, Heyrovsky, Hoar, Iannicola, Isasi~Parache,
  Janotto, Joliet, Jonckheere, Keil, Kim, Klagyivik, Klar, Knude, Kochukhov,
  Kolka, Kos, Kutka, Lainey, LeBouquin, Liu, Loreggia, Makarov, Marseille,
  Martayan, Martinez-Rubi, Massart, Meynadier, Mignot, Munari, Nguyen,
  Nordlander, Ocvirk, Olias~Sanz, Ortiz, Osorio, Oszkiewicz, Ouzounis, Palmer,
  Park, Pasquato, Peltzer, Peralta, P{\'{e}}turaud, Pieniluoma, \&
  Pigozzi}]{GaiaCollaboration2016TheMission}
{Gaia Collaboration}, Prusti, T., J~de Bruijne, J.~H., {et~al.} 2016, \aap,
  595, A1

\bibitem[{{Gaia Collaboration, Brown} {et~al.}(2018){Gaia Collaboration,
  Brown}, Vallenari, Prusti, de~Bruijne, Babusiaux, \&
  Bailer-Jones}]{GaiaCollaboration2018brown}
{Gaia Collaboration, Brown}, A. G.~A., Vallenari, A., Prusti, T., {et~al.}
  2018, \aap, 616, 21

\bibitem[{Grand {et~al.}(2019)Grand, Deason, White, Simpson, G{\'{o}}mez,
  Marinacci, \& Pakmor}]{Grand2019TheHalo}
Grand, R. J.~J., Deason, A.~J., White, S. D.~M., {et~al.} 2019, \mnras:
  Letters, 487, L72

\bibitem[{Grand {et~al.}(2017)Grand, G{\'{o}}mez, Marinacci, Pakmor, Springel,
  Campbell, Frenk, Jenkins, \& White}]{Grand2017TheTime}
Grand, R. J.~J., G{\'{o}}mez, F.~A., Marinacci, F., {et~al.} 2017, \mnras, 467,
  179

\bibitem[{Grand {et~al.}(2018)Grand, Helly, Fattahi, Cautun, Cole, Cooper,
  Deason, Frenk, G{\'{o}}mez, Hunt, Marinacci, Pakmor, Simpson, Springel, \&
  Xu}]{Grand2018Aurigaia:Simulations}
Grand, R. J.~J., Helly, J., Fattahi, A., {et~al.} 2018, \mnras, 481, 1726

\bibitem[{{GRAVITY Collaboration} {et~al.}(2018){GRAVITY Collaboration},
  Abuter, Amorim, Anugu, Baub{\"{o}}ck, Benisty, Berger, Blind, Bonnet,
  Brandner, Buron, Collin, Chapron, Cl{\'{e}}net, Coud{\'{e}} Du~Foresto,
  De~Zeeuw, Deen, Delplancke-Str{\"{o}}bele, Dembet, Dexter, Duvert, Eckart,
  Eisenhauer, Finger, F{\"{o}}rster~Schreiber, F{\'{e}}dou, Garcia,
  Garcia~Lopez, Gao, Gendron, Genzel, Gillessen, Gordo, Habibi, Haubois, Haug,
  Hau{\ss}mann, Henning, Hippler, Horrobin, Hubert, Hubin, Rosales, Jochum,
  Jocou, Kaufer, Kellner, Kendrew, Kervella, Kok, Kulas, Lacour,
  Lapeyr{\`{e}}re, Lazareff, Le~Bouquin, L{\'{e}}na, Lippa, Lenzen,
  M{\'{e}}rand, M{\"{u}}ler, Neumann, Ott, Palanca, Paumard, Pasquini, Perraut,
  Perrin, Pfuhl, Plewa, Rabien, Ram{\'{i}}rez, Ramos, Rau,
  Rodr{\'{i}}guez-Coira, Rohloff, Rousset, Sanchez-Bermudez, Scheithauer,
  Sch{\"{o}}ller, Schuler, Spyromilio, Straub, Straubmeier, Sturm, Tacconi,
  Tristram, Vincent, Von~Fellenberg, Wank, Waisberg, Widmann, Wieprecht, Wiest,
  Wiezorrek, Woillez, Yazici, Ziegler, \&
  Zins}]{GRAVITYCollaboration2018DetectionHole}
{GRAVITY Collaboration}, Abuter, R., Amorim, A., {et~al.} 2018, \aap, 615, 15

\bibitem[{Guillochon {et~al.}(2017)Guillochon, Parrent, Kelley, \&
  Margutti}]{Guillochon2017AnData}
Guillochon, J., Parrent, J., Kelley, L.~Z., \& Margutti, R. 2017, \apj, 835, 64

\bibitem[{Helmi {et~al.}(2018)Helmi, Babusiaux, Koppelman, Massari, Veljanoski,
  \& Brown}]{Helmi2018}
Helmi, A., Babusiaux, C., Koppelman, H.~H., {et~al.} 2018, Nature, 563, 85

\bibitem[{Hunt {et~al.}(2015)Hunt, Kawata, Grand, Minchev, Pasetto, \&
  Cropper}]{Hunt2015TheObservations}
Hunt, J. A.~S., Kawata, D., Grand, R. J.~J., {et~al.} 2015, \mnras, 450, 2132

\bibitem[{Hunter(2007)}]{Hunter2007Matplotlib:Environment}
Hunter, J.~D. 2007, Computing in Science {\&} Engineering, 9, 90

\bibitem[{Jaffe(1987)}]{Jaffe1987TheGalaxies}
Jaffe, W. 1987, Structure and Dynamics of Elliptical Galaxies, 127, 511

\bibitem[{Katz {et~al.}(2019)Katz, Sartoretti, Cropper, Panuzzo, Seabroke,
  Viala, Benson, Blomme, Jasniewicz, Jean-Antoine, Huckle, Smith, Baker, Crifo,
  Damerdji, David, Dolding, Fr{\'{e}}mat, Gosset, Guerrier, Guy, Haigron,
  Jan{\ss}en, Marchal, Plum, Soubiran, Th{\'{e}}venin, Ajaj, Allende~Prieto,
  Babusiaux, Boudreault, Chemin, Delle~Luche, Fabre, Gueguen, Hambly, Lasne,
  Meynadier, Pailler, Panem, Royer, Tauran, Zurbach, Zwitter, Arenou, Bossini,
  Gerssen, G{\'{o}}mez, Lemaitre, Leclerc, Morel, Munari, Turon, Vallenari, \&
  Erjal}]{Katz2019GaiaVelocities}
Katz, D., Sartoretti, P., Cropper, M., {et~al.} 2019, \aap, 622, 19

\bibitem[{Kluyver {et~al.}(2016)Kluyver, Ragan-Kelley, P{\'{e}}rez, Granger,
  Bussonnier, Frederic, Kelley, Hamrick, Grout, Corlay, Ivanov, Avila, Abdalla,
  Willing, \& Development~Team}]{Kluyver2016JupyterWorkflows}
Kluyver, T., Ragan-Kelley, B., P{\'{e}}rez, F., {et~al.} 2016, {Jupyter
  Notebooks-a publishing format for reproducible computational workflows} (IOS
  Press)

\bibitem[{Kochanek(1996)}]{Kochanek1996TheWay}
Kochanek, C.~S. 1996, \apj, 457, 228

\bibitem[{Koppelman \& Helmi(2021)}]{Koppelman2020TheCatalogue}
Koppelman, H.~H. \& Helmi, A. 2021, \aap, 645, A69

\bibitem[{Koppelman {et~al.}(2019)Koppelman, Helmi, Massari, Roelenga, \&
  Bastian}]{Koppelman2019a}
Koppelman, H.~H., Helmi, A., Massari, D., Roelenga, S., \& Bastian, U. 2019,
  \aap, 625, A5

\bibitem[{Kunder {et~al.}(2017)Kunder, Kordopatis, Steinmetz, Zwitter,
  Mcmillan, Casagrande, Enke, Wojno, Valentini, Chiappini, Matijevi{\v{c}},
  Siviero, De~Laverny, Recio-Blanco, Bijaoui, Wyse, Binney, Grebel, Helmi,
  Jofre, Antoja, Gilmore, Siebert, Famaey, Bienaym{\'{e}}, Gibson, Freeman,
  Miglio, \& Mosser}]{Kunder2017}
Kunder, A., Kordopatis, G., Steinmetz, M., {et~al.} 2017, \aj, 1, 12

\bibitem[{Leonard \& Tremaine(1990)}]{Leonard1990THESPEED}
Leonard, P. J.~T. \& Tremaine, S. 1990, \apj, 4, 486

\bibitem[{Leung \& Bovy(2019)}]{Leung2019SimultaneousLearning}
Leung, H.~W. \& Bovy, J. 2019, \mnras, 489, 2079

\bibitem[{Lindegren {et~al.}(2018)Lindegren, Hernandez, Bombrun, Klioner,
  Bastian, Ramos-Lerate, de~Torres, Steidelmuller, Stephenson, Hobbs, Lammers,
  Biermann, Geyer, Hilger, Michalik, Stampa, McMillan, Castaneda, Clotet,
  Comoretto, Davidson, Fabricius, Gracia, Hambly, Hutton, Mora, Portell, van
  Leeuwen, Abbas, Abreu, Altmann, Andrei, Anglada, Balaguer-Nunez, Barache,
  Becciani, Bertone, Bianchi, Bouquillon, Bourda, Brusemeister, Bucciarelli,
  Busonero, Buzzi, Cancelliere, Carlucci, Charlot, Cheek, Crosta, Crowley,
  de~Bruijne, de~Felice, Drimmel, Esquej, Fienga, Fraile, Gai, Garralda,
  Gonzalez-Vidal, Guerra, Hauser, Hofmann, Holl, Jordan, Lattanzi, Lenhardt,
  Liao, Licata, Lister, Loffler, Marchant, Martin-Fleitas, Messineo, Mignard,
  Morbidelli, Poggio, Riva, Rowell, Salguero, Sarasso, Sciacca, Siddiqui,
  Smart, Spagna, Steele, Taris, Torra, van Elteren, van Reeven, \&
  Vecchiato}]{Lindegren2018}
Lindegren, L., Hernandez, J., Bombrun, A., {et~al.} 2018, \aap, 616, A2

\bibitem[{Marchetti {et~al.}(2019)Marchetti, Rossi, \& Brown}]{Marchetti2018}
Marchetti, T., Rossi, E.~M., \& Brown, A. G.~A. 2019, \mnras, 490, 157

\bibitem[{Marrese {et~al.}(2019)Marrese, Marinoni, Fabrizio, \&
  Altavilla}]{Marrese2018}
Marrese, P.~M., Marinoni, S., Fabrizio, M., \& Altavilla, G. 2019, \aap, 621,
  A144

\bibitem[{McMillan(2017)}]{Mcmillan2017}
McMillan, P.~J. 2017, \mnras, 94, 76

\bibitem[{Monari {et~al.}(2018)Monari, Famaey, Carrillo, Piffl, Steinmetz,
  Wyse, Anders, Chiappini, \& Jan{\ss}en}]{Monari2018TheWay}
Monari, G., Famaey, B., Carrillo, I., {et~al.} 2018, \aap, 616, 9

\bibitem[{Navarro {et~al.}(1997)Navarro, Frenk, \&
  White}]{Navarro1997AClustering}
Navarro, J.~F., Frenk, C.~S., \& White, S. D.~M. 1997, \apj, 490, 493

\bibitem[{Piffl {et~al.}(2014{\natexlab{a}})Piffl, Binney, McMillan, Steinmetz,
  Helmi, Wyse, Bienayme, Bland-Hawthorn, Freeman, Gibson, Gilmore, Grebel,
  Kordopatis, Navarro, Parker, Reid, Seabroke, Siebert, Watson, \&
  Zwitter}]{Piffl2014}
Piffl, T., Binney, J., McMillan, P.~J., {et~al.} 2014{\natexlab{a}}, \mnras,
  445, 3133

\bibitem[{Piffl {et~al.}(2014{\natexlab{b}})Piffl, Scannapieco, Binney,
  Steinmetz, Scholz, Williams, de~Jong, Kordopatis, Matijevic, Bienayme,
  Bland-Hawthorn, Boeche, Freeman, Gibson, Gilmore, Grebel, Helmi, Munari,
  Navarro, Parker, Reid, Seabroke, Watson, Wyse, \& Zwitter}]{Piffl2014TheWay}
Piffl, T., Scannapieco, C., Binney, J., {et~al.} 2014{\natexlab{b}}, \aap, 562,
  A91

\bibitem[{Posti \& Helmi(2019)}]{Posti2019MassHubble}
Posti, L. \& Helmi, A. 2019, \aap, 621

\bibitem[{Schaye {et~al.}(2014)Schaye, Crain, Bower, Furlong, Schaller, Theuns,
  Dalla~Vecchia, Frenk, McCarthy, Helly, Jenkins, Rosas-Guevara, White, Baes,
  Booth, Camps, Navarro, Qu, Rahmati, Sawala, Thomas, \& Trayford}]{Schaye2014}
Schaye, J., Crain, R.~A., Bower, R.~G., {et~al.} 2014, \mnras, 446, 521

\bibitem[{Sch{\"{o}}nrich {et~al.}(2010)Sch{\"{o}}nrich, Binney, \&
  Dehnen}]{Schonrich2010}
Sch{\"{o}}nrich, R., Binney, J., \& Dehnen, W. 2010, \mnras, 403, 1829

\bibitem[{Sch{\"{o}}nrich {et~al.}(2019)Sch{\"{o}}nrich, Mcmillan, \&
  Eyer}]{Schonrich2019}
Sch{\"{o}}nrich, R., Mcmillan, P., \& Eyer, L. 2019, \mnras, 487, 3568

\bibitem[{Smith {et~al.}(2007)Smith, Ruchti, Helmi, Wyse, Fulbright, Freeman,
  Navarro, Seabroke, Steinmetz, Williams, Bienaym, Binney, Dehnen, Gibson,
  Gilmore, Grebel, Munari, Parker, Scholz, Siebert, Watson, \&
  Zwitter}]{Smith2007}
Smith, M.~C., Ruchti, G.~R., Helmi, A., {et~al.} 2007, \mnras, 772, 755

\bibitem[{Steinmetz {et~al.}(2006)Steinmetz, Zwitter, Siebert, Watson, Freeman,
  Munari, Campbell, Williams, Seabroke, Wyse, Parker, Bienaym{\'{e}}, Roeser,
  Gibson, Gilmore, Grebel, Helmi, Navarro, Burton, Cass, Dawe, Fiegert,
  Hartley, Russell, Saunders, Enke, Bailin, Binney, Bland-Hawthorn, Boeche,
  Dehnen, Eisenstein, Evans, Fiorucci, Fulbright, Gerhard, Jauregi, Kelz,
  Mijovic´, Minchev, Parmentier, Pe{\~{n}}arrubia, Quillen, Read, Ruchti,
  Scholz, Siviero, Smith, Sordo, Veltz, Vidrih, Von~Berlepsch, Boyle, \&
  Schilbach}]{Steinmetz2006THERELEASE}
Steinmetz, M., Zwitter, T., Siebert, A., {et~al.} 2006, \aj, 132, 1645

\bibitem[{Tremaine(1987)}]{Tremaine1987Summary}
Tremaine, S. 1987, Structure and Dynamics of Elliptical Galaxies, 127, 367

\bibitem[{Van Der~Walt {et~al.}(2011)Van Der~Walt, Colbert, \&
  Varoquaux}]{VanDerWalt2011TheComputation}
Van Der~Walt, S., Colbert, S.~C., \& Varoquaux, G. 2011, Computing in Science
  and Engineering, 13, 22

\bibitem[{Vasiliev(2019)}]{Vasiliev2019}
Vasiliev, E. 2019, \mnras, 482, 1525

\bibitem[{Virtanen {et~al.}(2020)Virtanen, Gommers, Oliphant, Haberland, Reddy,
  Cournapeau, Burovski, Peterson, Weckesser, Bright,
  {et~al.}}]{Virtanen2020SciPyPython}
Virtanen, P., Gommers, R., Oliphant, T.~E., {et~al.} 2020, Nature Methods, 17,
  261

\bibitem[{Watkins {et~al.}(2010)Watkins, Evans, \& An}]{Watkins2010TheGalaxies}
Watkins, L.~L., Evans, N.~W., \& An, J.~H. 2010, \mnras, 406, 264

\bibitem[{Williams {et~al.}(2017)Williams, Belokurov, Casey, \&
  Evans}]{Williams2017}
Williams, A.~A., Belokurov, V., Casey, A.~R., \& Evans, N.~W. 2017, \mnras,
  468, 2359

\bibitem[{Wilson {et~al.}(2010)Wilson, Hearty, Skrutskie, Majewski, Schiavon,
  Eisenstein, Gunn, Blank, Henderson, Smee, Barkhouser, Harding, Fitzgerald,
  Stolberg, Arns, Nelson, Brunner, Burton, Walker, Lam, Maseman, Barr, Leger,
  Carey, MacDonald, Horne, Young, Rieke, Rieke, O'Brien, Hope, Krakula, Crane,
  Zhao, Carr, Harrison, Stoll, Vernieri, Holtzman, Shetrone, Allende-Prieto,
  Johnson, Frinchaboy, Zasowski, Bizyaev, Gillespie, \&
  Weinberg}]{Wilson2010TheSpectrograph}
Wilson, J.~C., Hearty, F., Skrutskie, M.~F., {et~al.} 2010, in Ground-based and
  Airborne Instrumentation for Astronomy III, ed. I.~S. McLean, S.~K. Ramsay,
  \& H.~Takami, Vol. 7735 (International Society for Optics and Photonics),
  77351C

\bibitem[{{Wolfram Research{,} Inc.}(2020)}]{Mathematica}
{Wolfram Research{,} Inc.} 2020, Mathematica, {V}ersion 12.1, champaign, IL,
  2020

\bibitem[{Xue {et~al.}(2008)Xue, Rix, Zhao, Re~Fiorentin, Naab, Steinmetz,
  van~den Bosch, Beers, Lee, Bell, Rockosi, Yanny, Newberg, Wilhelm, Kang,
  Smith, \& Schneider}]{Xue2008}
Xue, X.~X., Rix, H.~W., Zhao, G., {et~al.} 2008, \apj, 684, 1143

\bibitem[{Zaritsky {et~al.}(2020)Zaritsky, Conroy, Zhang, Bonaca, Caldwell,
  Cargile, Johnson, \& Naidu}]{Zaritsky2019ASurvey}
Zaritsky, D., Conroy, C., Zhang, H., {et~al.} 2020, \apj, 888, 114

\bibitem[{Zinn {et~al.}(2019)Zinn, Pinsonneault, Huber, \&
  Stello}]{Zinn2019ConfirmationField}
Zinn, J.~C., Pinsonneault, M.~H., Huber, D., \& Stello, D. 2019, \apj, 878, 136

\end{thebibliography}


\end{document}